\documentclass[12pt]{article}

\usepackage{amsmath}
\usepackage{amssymb}
\usepackage{latexsym}
\usepackage{graphicx}
\usepackage{epsfig}

\def\beq{\begin{equation}}
\def\eeq{\end{equation}}
\def\bea{\begin{eqnarray}}
\def\eea{\end{eqnarray}}

\begin{document}

\begin{titlepage}

\vspace*{1cm}
\begin{center}
{\bf \Large Effective Temperatures and Radiation Spectra for\\[2mm]
a Higher-Dimensional Schwarzschild-de-Sitter\\[4mm] Black-Hole}

\bigskip \bigskip \medskip

{\bf P. Kanti} and {\bf T. Pappas}

\bigskip
{\it Division of Theoretical Physics, Department of Physics,\\
University of Ioannina, Ioannina GR-45110, Greece}

\bigskip \medskip
{\bf Abstract}
\end{center}
The absence of a true thermodynamical equilibrium for an observer located in
the causal area of a Schwarzschild-de Sitter spacetime has repeatedly raised 
the question of the correct definition of its temperature. In this work, we
consider five different temperatures for a higher-dimensional 
Schwarzschild-de Sitter black hole: the bare $T_0$, the normalised $T_{BH}$
and three effective ones given in terms of both the black hole and cosmological
horizon temperatures. We find that these five temperatures exhibit similarities but
also significant differences in their behaviour as the number of extra dimensions
and the value of the cosmological constant are varied. We then investigate their 
effect on the energy emission spectra of Hawking radiation. We demonstrate that the
radiation spectra for the normalised temperature $T_{BH}$ -- proposed by Bousso
and Hawking over twenty years ago -- leads to the dominant emission curve
while the other temperatures either support a significant emission rate only
at a specific $\Lambda$ regime or they have their emission rates globally
suppressed. Finally, we compute the bulk-over-brane emissivity ratio and 
show that the use of different temperatures may lead to different conclusions
regarding the brane or bulk dominance.

\end{titlepage}

\setcounter{page}{1}

\section{Introduction}

The novel theories, that postulate the existence of additional spacelike dimensions in
nature \cite{ADD,RS} with size much larger than the Planck length or even infinite, have
in fact an almost 20-year life-time. During that period, several aspects of gravity, cosmology
and particle physics have been reconsidered in the context of these higher-dimensional
theories. Black hole solutions have been intensively studied since the existence of extra
dimensions affects both their creation and decay processes (for more information on this,
one may consult the reviews \cite{Kanti:2004}-\cite{PKEW}). 

The presence of the brane(s) in the model with warped extra dimensions \cite{RS} has
proven so far to be an unsurmountable obstacle for the construction of analytical
solutions describing regular black holes. As a result, most of the study of the decay
process of a higher-dimensional black hole has been restricted in the context of the
model with large extra dimensions \cite{ADD}, where the latter are assumed
to be empty, and thus flat, and where the self-energy of the brane may be ignored
compared to the black-hole mass.  It is in the context of this theory that analytical
expressions describing higher-dimensional black holes may be written, and the 
emission of particles, comprising the Hawking radiation \cite{Hawking}, may be studied
in detail. 

Historically, the first solution describing a higher-dimensional, spherically-symmetric
black hole appeared in the '60s, and is known as the Tangherlini solution \cite{Tangherlini}.
The solution describes a higher-dimensional analogue of the Schwarzschild solution 
of the General Theory of Relativity that is formed also in the presence of a cosmological
constant. Therefore, this solution constitutes in fact an improvement of the assumption
made in the context of the large extra dimensions scenario where the extra space is
absolutely empty: here, the extra dimensions are filled with a constant distribution of
energy, or with some field configuration that effectively acts as a constant
distribution of energy. For a positive cosmological constant, the solution describes a
higher-dimensional Schwarzschild-de Sitter black-hole spacetime.

Although the emission of Hawking radiation from higher-dimensional, spherically-symmetric
or rotating black holes has been extensively studied in the literature (for a partial only
list see \cite{KMR}-\cite{Miao} or the aforementioned reviews \cite{Kanti:2004}-\cite{PKEW}),
the analyses focused on the higher-dimensional Schwarzschild-de Sitter black holes
are only a few. The first such work \cite{KGB} contained an analytic study of the 
greybody factor for scalar fields propagating on the brane and in the bulk, and in
addition provided exact numerical results for the radiation spectra in both emission
channels. A subsequent analytic work \cite{Harmark} extended the aforementioned
analysis by determining the next-to-leading-order term in the expansion of the
greybody factor. An exact numerical study \cite{Wu} then considered the emission
of fields with arbitrary spin from a higher-dimensional Schwarzschild-de-Sitter
black hole. A series of three, more recent works studied the case of a scalar field
having a non-minimal coupling to the scalar curvature: the first \cite{Crispino}
studied the case of a purely 4-dimensional Schwarzschild-de-Sitter black hole,
the second \cite{KPP1} considered the scalar field propagating either in the
higher-dimensional bulk or being restricted on a brane, and a third one \cite{KPP3}
provided exact numerical results for the greybody factors and radiation spectra
in the same theory. A few additional works 
\cite{Anderson, Sporea, Ahmed, Fernando, Boonserm} have also appeared
that studied the greybody factors for fields propagating in the
background of variants of a Schwarzschild-de-Sitter black hole.

However, over the years, the question of what is the correct notion of the
temperature of a Schwarzschild - de Sitter (SdS) spacetime has risen. This spacetime
contains a black hole whose event horizon sets the lower boundary of the causally
connected spacetime. But it also contains a positive cosmological constant that gives
rise to a cosmological horizon, the upper boundary of the causal spacetime. An
observer living at any point of this causal area is never in a true thermodynamical
equilibrium - the two horizons have each one its own temperature, expressed in
terms of their surface gravities \cite{GH1, GH2}, and thus an incessant flow of thermal
energy (from the hotter black-hole horizon to the colder cosmological one) takes
place at every moment. In addition, the SdS spacetime lacks an asymptotically-flat
limit where the black-hole parameters may be defined in a robust way. The latter
problem was solved in \cite{BH} where a {\it normalised} black-hole temperature was
proposed that made amends for the lack of an asymptotic limit. Then, assuming
that the value of the cosmological constant is small and the two horizons are thus
located far away from each other, one could formulate two independent thermodynamics.

Despite the above, the question of what happens as the cosmological constant becomes larger
and the two horizons come closer still persisted. It was this question that gave rise to
the notion of the {\it effective temperature} for an SdS spacetime 
\cite{Shankar, Urano, Lahiri, Bhatta, LiMa}, namely one that implements both the
black-hole and the cosmological horizon temperatures (for a review on this, see
\cite{Mann_review}). A number of additional works have appeared in the literature
with similar or alternative approaches on the thermodynamics of de Sitter spacetimes
\cite{Romans}-\cite{Pourhassan}, however, the question of the appropriate expression
of the SdS black-hole temperature still remains open.

Up to now, no work has appeared in the literature that makes a comprehensive study
of the different temperatures for an SdS spacetime and compare their predictions for 
the corresponding Hawking radiation spectra. In fact, previous works that study the
radiation spectra from a four-dimensional or higher-dimensional SdS black hole make
use of either its {\it bare} temperature $T_0$, based on its surface gravity, or the
normalised one $T_{BH}$, at will. In the context of this work, we will perform such a
comprehensive study, and we will derive and compare the derived radiation spectra.
We will do so not only for the aforementioned two SdS black-hole temperatures
but also for three additional effective temperatures for the SdS spacetime, namely $T_{eff-}$,
$T_{eff+}$ and $T_{effBH}$ -- the use of one of the latter temperatures may be
unavoidable for large values of the cosmological constant when the two horizons lie
so close that the independent thermodynamics no longer hold. To address the above,
we will also extend the regime of values of the cosmological constant that has been
studied in the literature so far, and consider the entire allowed regime, from a very
small value up to its maximum critical value \cite{Nariai}. 

To make our analysis as general as possible, we will consider a higher-dimensional
SdS spacetime. We will then study the properties of the different temperatures both
in terms of the value of the cosmological constant but also of the number of extra
spacelike dimensions. The corresponding Hawking radiation spectra will then be produced
for scalar fields, both minimally and non-minimally coupled to gravity, propagating either
on our brane or in the bulk.  As we will see, the different temperatures will lead to
different energy emission rates for the black hole, each one with its own profile in
terms of the bulk cosmological constant, number of extra dimensions and value of
the non-minimal coupling constant. In addition, each temperature will lead to different
conclusions regarding the dominance of the brane or of the bulk.

The outline of our paper is as follows: in Section 2, we present the theoretical
framework of our analysis, the gravitational background, the equations of motion
for the scalar field as well as the different definitions of the temperature of an
SdS spacetime. In Sections 3 and 4, we derive the energy emission rates for bulk
and brane scalar fields, having a minimal or non-minimal coupling to gravity,
respectively. In Section 5, we calculate the bulk-over-brane emissivity ratio and,
in Section 6, we summarise our analysis and present our conclusions.


\section{The Theoretical Framework}

\subsection{The Gravitational Background}

We will start by considering a higher-dimensional gravitational theory with $D=4+n$
total number of dimensions. The action functional of the theory will also contain a positive
cosmological constant $\Lambda$, and will therefore read 
\beq
S_D=\int d^{4+n}x\, \sqrt{-G}\,\left(\frac{R_D}{2 \kappa^2_D} - \Lambda\right)\,.
\label{action_D}
\eeq
In the above, $R_D$ is the higher-dimensional Ricci scalar and $\kappa^2_D=1/M_*^{2+n}$
the higher-dimensional gravitational constant associated with the fundamental scale of gravity $M_*$.
If we vary the above action with respect to the metric tensor $G_{MN}$, we obtain the 
Einstein's field equations that have the form
\beq
R_{MN}-\frac{1}{2}\,G_{MN}\,R_D = \kappa^2_D\,T_{MN} = -\kappa^2_D G_{MN} \Lambda\,,
\label{field_eqs}
\eeq
with the only contribution to the energy-momentum tensor $T_{MN}$ coming from the
bulk cosmological constant. 

The above set of equations admit a spherically-symmetric solution of the form
\cite{Tangherlini}
\beq
ds^2 = - h(r)\,dt^2 + \frac{dr^2}{h(r)} + r^2 d\Omega_{2+n}^2,
\label{bhmetric}
\eeq
where $d\Omega_{2+n}^2$ is the area of the ($2+n$)-dimensional unit sphere
given by
\begin{equation}
d\Omega_{2+n}^2=d\theta^2_{n+1} + \sin^2\theta_{n+1} \,\biggl(d\theta_n^2 +
\sin^2\theta_n\,\Bigl(\,... + \sin^2\theta_2\,(d\theta_1^2 + \sin^2 \theta_1
\,d\varphi^2)\,...\,\Bigr)\biggr)\,,
\label{unit}
\end{equation}
with $0 \leq \varphi < 2 \pi$ and $0 \leq \theta_i \leq \pi$, for $i=1, ..., n+1$. The
radial function $h(r)$ is found to have the explicit form \cite{Tangherlini}
\beq
h(r) = 1-\frac{\mu}{r^{n+1}} - \frac{2 \kappa_D^2\,\Lambda r^2}{(n+3) (n+2)}\,.
\label{h-fun}
\eeq
The above gravitational background describes a ($4+n$)-dimensional Schwarzschild-de-Sitter
(SdS) spacetime, with the parameter $\mu$ related to the black-hole mass $M$ through the
relation \cite{MP}
\beq
\mu=\frac{\kappa^2_D M}{(n+2)}\,\frac{\Gamma[(n+3)/2]}{\pi^{(n+3)/2}}\,.
\eeq
The horizons of the SdS black hole follows from the equation $h(r)=0$ -- this has, in principle,
$(n+3)$ roots, however, not all of them are real and positive; in fact, the SdS spacetime
may have two, one or zero horizons, depending on the values of the parameters $M$ and
$\Lambda$ \cite{Molina}. Here, we will ensure that the values of $M$ and $\Lambda$
are in the regime that supports the existence of two horizons, the black-hole $r_h$ and the
cosmological one $r_c$, with $r_h<r_c$. However, the degenerate case, that results in the
Nariai limit \cite{Nariai} in which the two horizons coincide, will also be investigated. 

The higher-dimensional background (\ref{bhmetric}) is seen by gravitons and particles
with no Standard-Model quantum numbers that may propagate in the bulk. All ordinary
particles, however, are restricted to live on our 4-dimensional brane \cite{ADD,RS},
and therefore propagate on a different gravitational background. The latter follows by 
projecting the $(4+n)$-dimensional background (\ref{bhmetric}) on the brane and it
is realised by fixing the value of the extra angular coordinates, $\theta_i=\pi/2$,
for $i=2, ..., n+1$. Then, we obtain the 4D line-element
\beq
ds^2 = - h(r)\,dt^2 + \frac{dr^2}{h(r)} + r^2\,(d\theta^2 + \sin^2\theta\,
d\varphi^2)\,,
\label{metric_brane}
\eeq
with the metric function $h(r)$ preserving its form, given by Eq. (\ref{h-fun}), and
thus its dependence on both the number of additional spacelike coordinates $n$ and
the value of the bulk cosmological constant $\Lambda$.


\subsection{The Temperature of the Schwarzschild-de Sitter Black Hole} 

The temperature of a black hole is traditionally defined in terms of its surface
gravity $k_h$ at the location of the horizon \cite{GH1, GH2}. The latter quantity is
expressed as
\beq
k_h^2 =-\frac{1}{2}\,\lim_{r \rightarrow r_h} (D_M K_N)(D^M K^N)\,,
\label{surface_grav1}
\eeq
where $D_M$ is the covariant derivative and 
\beq
K=\gamma_t\,\frac{\partial \,\,}{\partial t}
\eeq
is the timelike Killing vector with $\gamma_t$ a normalization constant. In the case
that the gravitational background is spherically-symmetric, Eq. (\ref{surface_grav1})
takes the simpler form \cite{York}
\beq
k_h=\frac{1}{2}\,\frac{1}{\sqrt{-g_{tt} g_{rr}}}\,|g_{tt,r}|_{r=r_h}\,.
\label{surface_grav2}
\eeq
When the above expression is employed for the line-element (\ref{bhmetric}) of a
higher-dimensional Schwarzschild-de Sitter black hole, we obtain the following
expression for its temperature \cite{GH1, York, KGB}
\beq
T_0 =  \frac{k_h}{2\pi}=\frac{1}{4\pi r_h}\,\left[(n+1)-(n+3) \tilde \Lambda r_h^2\right],
\label{Temp0}
\eeq
where we have defined, for convenience, the quantity 
$\tilde \Lambda=2 \kappa^2_D \Lambda/(n+2)(n+3)$, and used the condition
$f(r_h)=0$ to replace $\mu$ in terms of $r_h$ and $\tilde \Lambda$.

The Schwarzschild-de Sitter spacetime is characterized, in the most generic case, by the
presence of a second horizon, the cosmological horizon $r_c$. As a result, one may define
another surface gravity $k_c$, this time at the location of $r_c$, and a
temperature for the cosmological horizon \cite{GH1, GH2}, namely \cite{KGB}
\beq
T_c = - \frac{k_c}{2\pi}=-\frac{1}{4\pi r_c}\,\left[(n+1)-(n+3) \tilde \Lambda r_c^2\right],
\label{Tempc}
\eeq
where care has been taken so that $T_c$ is positive-definite since $r_h<r_c$ \cite{KGB}.
The presence of the second horizon with its own temperature makes the thermodynamics
of the Schwarzschild-de Sitter spacetime significantly more complicated, as compared
to the cases of either asymptotically Minkowski or Anti de Sitter spacetimes
\cite{Mann_review}. The two temperatures, $T_0$ and $T_c$, are in principle different,
therefore an observer located at an arbitrary point of the causal
region $r_h<r<r_c$ is not in thermodynamical equilibrium. The usual approach adopted
in the literature is to make the assumption that the two horizons are located far away and 
therefore each one can have its own independent thermodynamics \cite{GH2, BH, Sekiwa} -- this
assumption, however, is valid only for small values of the cosmological constant and thus
it imposes a constraint on all potential analyses.

In \cite{BH}, a modified expression for the temperature of the black hole was proposed,
namely
\beq
T_{BH} = \frac{1}{\sqrt{h(r_0)}}\,\frac{1}{4\pi r_h}\,\left[(n+1)-(n+3) \tilde \Lambda r_h^2\right],
\label{TempBH}
\eeq
in which a normalization factor $\sqrt{h(r_0)}$ was introduced involving the value of the
metric function at its global maximum $r_0$. This point follows from the condition
$h'(r)=0$ and is given by \cite{KGB}
\beq 
r_0^{n+3}=\frac{(n+1) \mu}{2 \tilde \Lambda}\,.
\label{def-r0}
\eeq
There, the metric function assumes the value
\beq
h(r_0)=1-\frac{\mu}{r_0^{n+1}} -\tilde \Lambda r_0^2=\frac{1}{n+1}\,
\left[(n+1) - (n+3) \tilde \Lambda r_0^2\right].
\label{h_r0}
\eeq
The above is the maximum value that the metric function attains as it interpolates between
the two zeros at the two horizons. The point $r_0$ is the point the closest that the
Schwarzschild-de Sitter spacetime has to an asymptotically flat region: it is here that the
effects of the black-hole and cosmological horizons cancel out and an observer can stay
at rest \cite{BH}. Mathematically, the normalization factor $\sqrt{h(r_0)}$ appears from the
normalization of the Killing vector, $K_M K^M=-1$: this condition is satisfied in asymptotically
flat spacetime for $\gamma_t=1$ but, at $r=r_0$, this factor should be
$\gamma_t=1/\sqrt{h(r_0)}$. 

\begin{figure}[t]
  \begin{center}
\mbox{\includegraphics[width = 0.50 \textwidth] {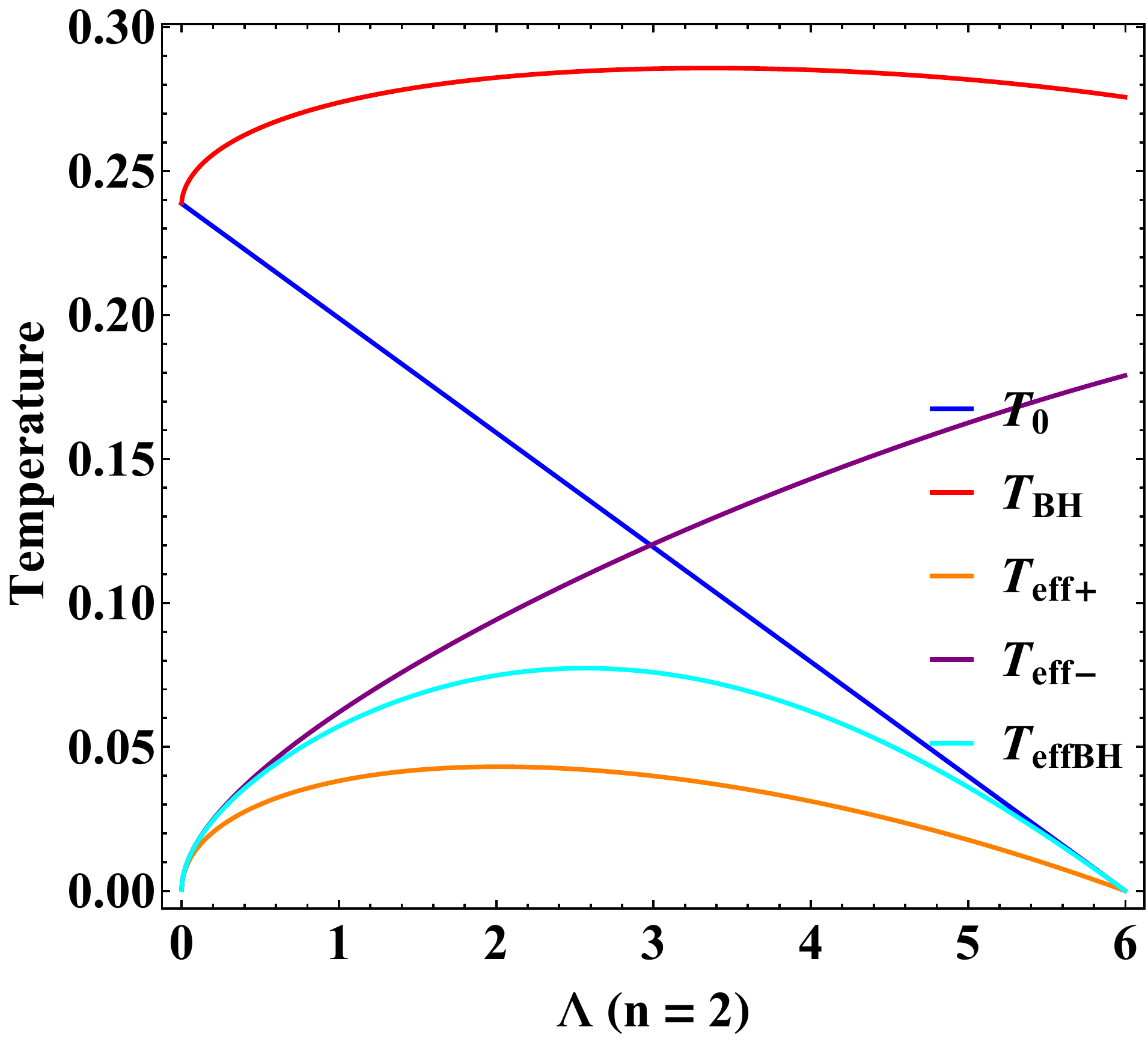}}
\hspace*{0.3cm} {\includegraphics[width = 0.41 \textwidth]
{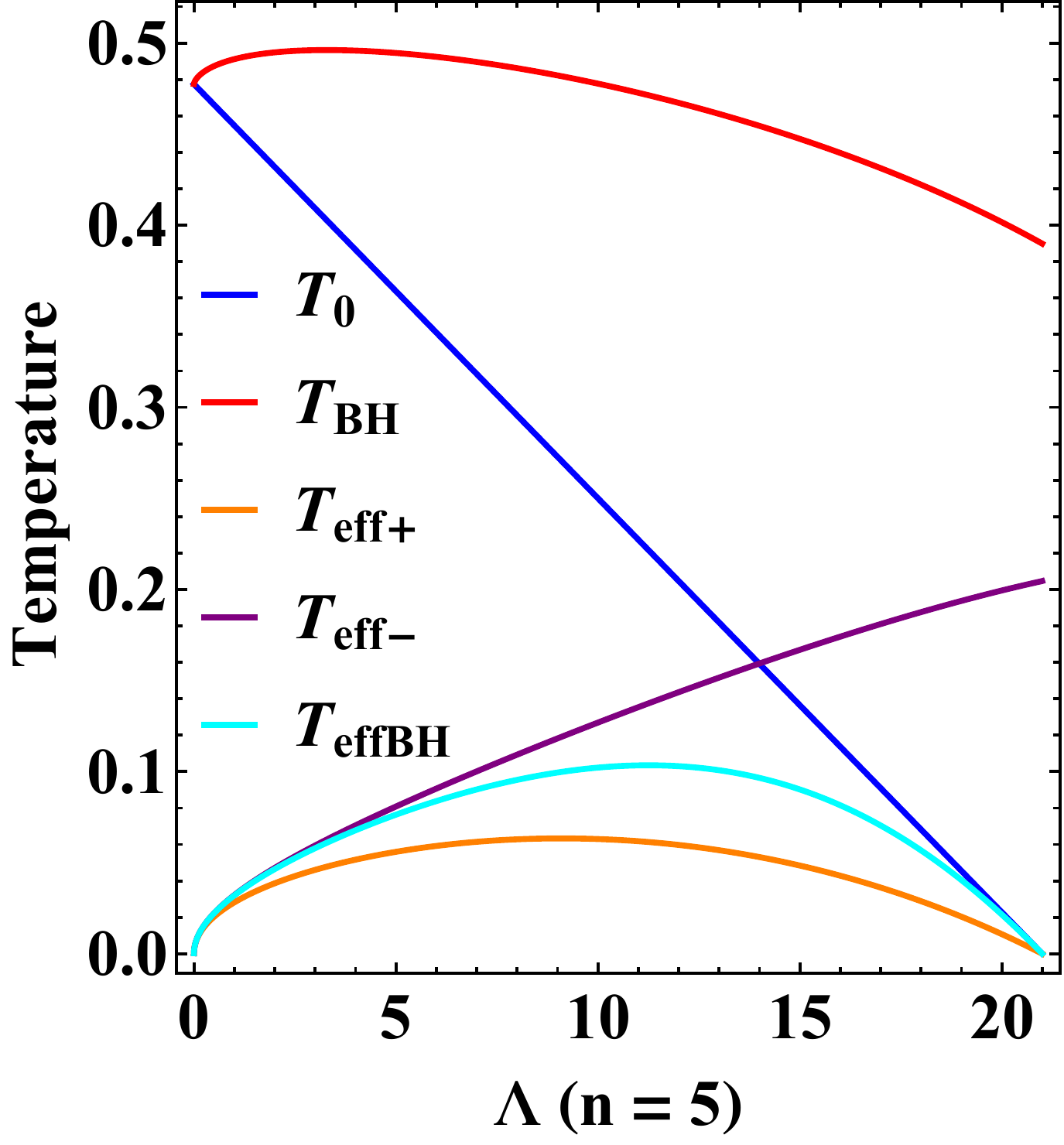}}
    \caption{Temperatures for a $(4+n)$-dimensional Schwarzschild-de Sitter
    black hole as a function of the cosmological constant $\Lambda$, for:
    \textbf{(a)} $n=2$, and \textbf{(b)} $n=5$.}
   \label{Temp_L_n1}
  \end{center}
\end{figure}

Including this normalisation factor in Eq. (\ref{TempBH}) is a step forward in defining
the black-hole temperature in a non-asymptotically flat spacetime, however, 
this factor modifies significantly the properties of $T_0$. In Fig. \ref{Temp_L_n1}(a,b),
we depict the dependence of the two temperatures, $T_0$ and $T_{BH}$, as a function of
the cosmological constant, and for two values of the number of extra dimensions,
$n=2$ and $n=5$. For low $n$, as $\Lambda$ increases, $T_0$ monotonically decreases,
in accordance to Eq. (\ref{Temp0}), whereas $T_{BH}$ predominantly increases - the latter
is caused by the variation in the value of $h(r_0)$ that, in most part of the allowed 
$\Lambda$ regime, causes an enhancement in $T_{BH}$. For large
values of $n$, the monotonic decrease of $T_0$ remains unaffected while the increase of
$T_{BH}$ holds only for the lower range of values of $\Lambda$. Even in this case, 
the value of $T_{BH}$ is constantly larger than that of $T_0$ (see, also, \cite{Wu}
for a similar comparison and conclusions). The two temperatures
match only in the limit $\Lambda \rightarrow 0$ when they reduce to the temperature
of a higher-dimensional Schwarzschild black hole. A radically different behaviour appears
in the opposite limit, the Nariai or extremal limit \cite{Romans, Traschen, Nariai}:
as $\Lambda$
approaches its maximum allowed value, the two horizons approach each other and eventually
coincide, with $r_h=r_c$. In that limit, the combination inside the square brackets in
Eq. (\ref{Temp0}), and thus $T_0$ itself, vanishes\footnote{Although, for arbitrary $n$, this
is very difficult to prove analytically, for special values of $n$ we may easily confirm it: for
$n=0$, the Nariai limit is reached when $M^2 \Lambda=1/9$ and then $r_h^2 =1/\Lambda$;
for $n=1$, the two horizons coincide when $\mu \tilde \Lambda=1/4$ and then
$r_h^2=1/(2\tilde \Lambda)$. In both cases, we may easily see that Eq. (\ref{Temp0})
vanishes. For higher values of $n$, the vanishing of Eq. (\ref{Temp0}) may be easily
confirmed numerically.},
a feature that is clearly shown in Fig. \ref{Temp_L_n1}. On the contrary,
in the critical limit, $T_{BH}$ assumes an asymptotic constant value; this is caused by
the fact that its numerator and denominator both tend to zero values with the
ratio approaching a constant number.

\begin{figure}[t]
  \begin{center}
\mbox{\includegraphics[width = 0.43 \textwidth] {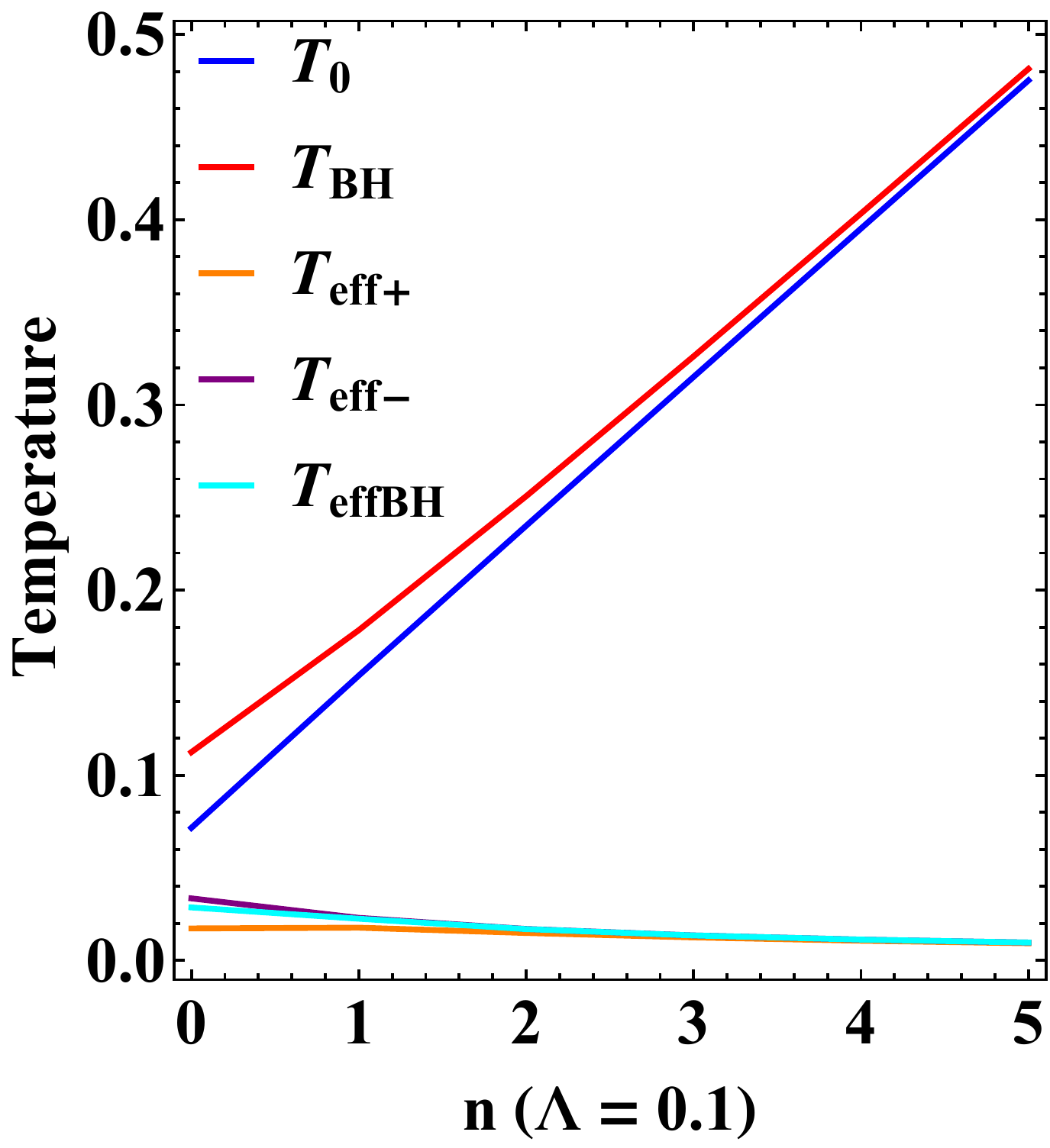}}
\hspace*{0.3cm} {\includegraphics[width = 0.43 \textwidth]
{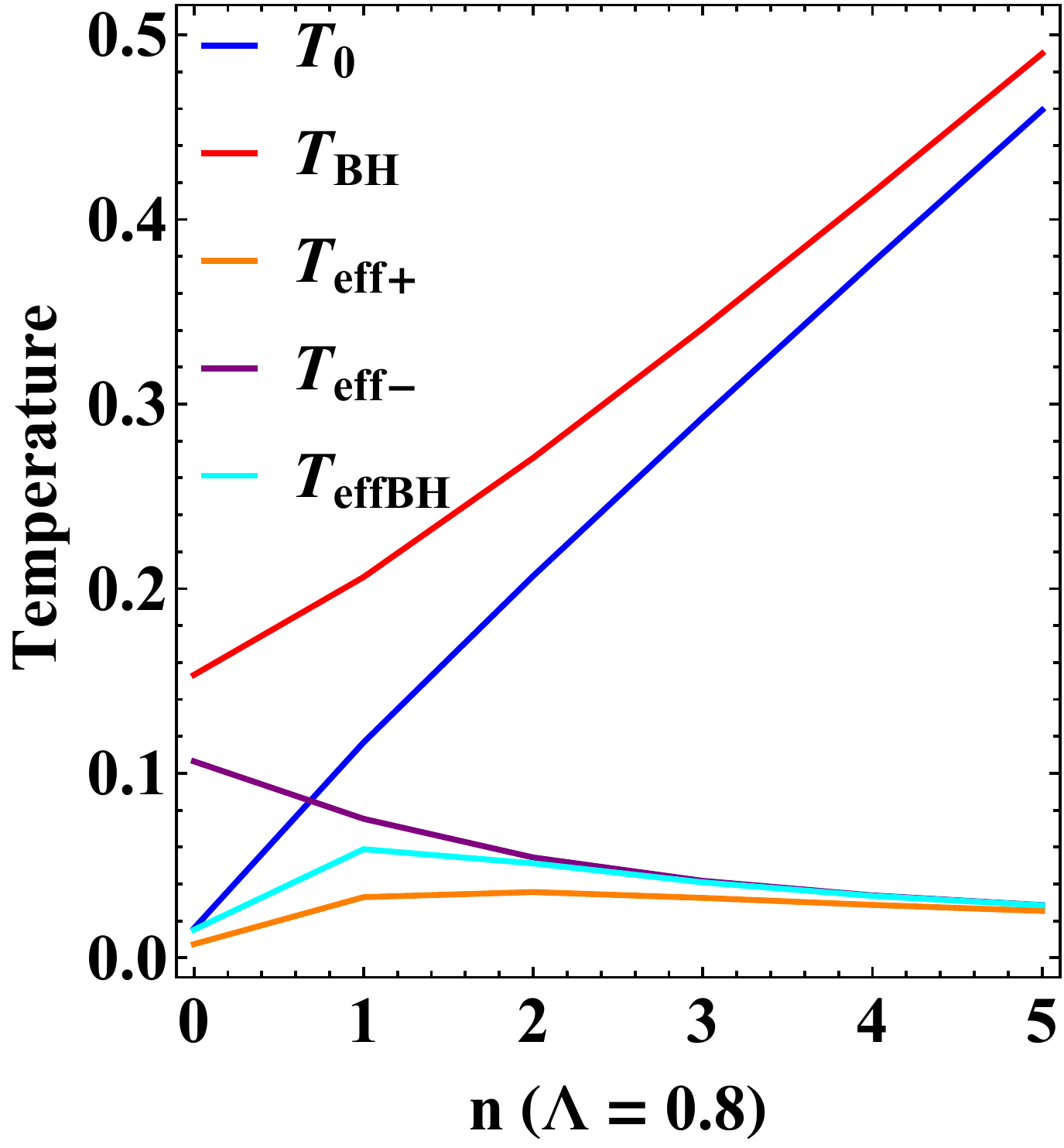}}
    \caption{Temperatures for a $(4+n)$-dimensional Schwarzschild-de Sitter
    black hole as a function of the number of extra dimensions $n$, for:
    \textbf{(a)} $\Lambda=0.1$, and \textbf{(b)} $\Lambda=0.8$.}
   \label{Temp_n_L08}
  \end{center}
\end{figure}

In Fig. \ref{Temp_n_L08}, we show the dependence of $T_0$ and $T_{BH}$ on
the number of extra dimensions $n$, for two different fixed values of the 
cosmological constant, $\Lambda=0.1$ and $\Lambda=0.8$ (we have set for simplicity
$\kappa_D^2=1$, therefore $\Lambda$ is given in units of
$r_h^{-2}$). We observe again that the `normalised' temperature $T_{BH}$ remains
always larger than the `bare' one $T_0$, however this dominance gets softer as
$n$ increases, and almost disappears for small values of $\Lambda$.

The temperature of a black hole is one of the important factors that determine the 
Hawking radiation emission spectra. Only a handful of works exist in the literature that
study the emission of Hawking radiation from a Schwarzschild-de Sitter black hole,
either 4-dimensional or higher-dimensional, and these use both definitions of its
temperature, Eq. (\ref{Temp0}) \cite{Crispino, Labbe} or Eq. (\ref{TempBH}) 
\cite{KGB, Wu, KPP3}, at will.
In addition, during the recent years, the notion of the {\it effective temperature} of
the Schwarzschild-de Sitter spacetime has emerged, that involves both temperatures
$T_0$ and $T_c$, in an attempt to unify the thermodynamical description of this
spacetime. In the most popular of the analyses, a thermodynamical first law for
a Schwarzschild-de Sitter black hole is written in which the black-hole mass 
plays the role of the enthalpy of the system ($M=-H$), the cosmological
constant that of the pressure ($P=\Lambda/8\pi$) while the entropy is the sum
of the entropies of the two horizons ($S=S_h+S_c$) \cite{Shankar, Urano, Lahiri, Bhatta,
LiMa, Mann_review}. In this picture, an effective temperature emerges that has the form
\beq
T_{eff-}=\left(\frac{1}{T_c}-\frac{1}{T_0}\right)^{-1}=\frac{T_0 T_c}{T_0-T_c}\,.
\label{Teff-}
\eeq
The above expression was obtained for the case of a 4-dimensional Schwarzschild-de Sitter
black hole. However, the arguments leading to the formulation of the aforementioned first
thermodynamical law had no explicit dependence on the dimensionality of spacetime. 
Therefore, we expect that the functional form of the effective temperature $T_{eff-}$
for the case of a $(4+n)$-dimensional Schwarzschild-de Sitter black hole will still
be given by Eq. (\ref{Teff-}), but with the individual temperatures $T_0$ and $T_c$,
now assuming their higher-dimensional forms, Eqs. (\ref{Temp0}) and (\ref{Tempc}). 
Then, the explicit form of $T_{eff-}$ in $D=4+n$ dimensions will be the following
\beq
T_{eff-}=-\frac{1}{4\pi}\,\frac{(n+1)^2 -(n+1)(n+3) \tilde \Lambda (r_h^2+r_c^2)
+(n+3)^2 \tilde \Lambda^2 r_h^2 r_c^2}{(r_h +r_c)\,[(n+1) -(n+3) \tilde \Lambda r_h r_c]}\,.
\label{Teff-expl}
\eeq
In the limit $r_h \rightarrow 0$, the above expression for the effective temperature reduces to 
that of the cosmological horizon $T_c$, as expected. However, the limit $r_c \rightarrow \infty$
(or, equivalently, $\tilde \Lambda \rightarrow 0$) leads to a vanishing result: the effective
temperature does not interpolate between the black-hole temperature $T_0$ and the
cosmological one $T_c$, as one may have expected; in fact, the limit $\Lambda 
\rightarrow 0$ is a particular one since it is equivalent to a vanishing pressure of the
system, that in the relevant analyses is always assumed to be positive. That is, by
construction, $T_{eff-}$ is valid for non-vanishing cosmological constant - but this is
exactly the regime where the need for an effective temperature really emerges since, in
the limit of small $\Lambda$, the horizons $r_h$ and $r_c$ are located so
far away from each other that the independent thermodynamics at the two horizons do
indeed hold.

In Figs. \ref{Temp_L_n1} and \ref{Temp_n_L08}, we depict also the behaviour of $T_{eff-}$
in terms of the value of the cosmological constant $\Lambda$ and the number of extra
dimensions $n$, respectively. The effective temperature $T_{eff-}$ is an increasing function
of $\Lambda$, and, similarly to the case of the normalised temperature $T_{BH}$, it assumes
a non-vanishing constant value at the critical limit - as in the case of $T_{BH}$, the
numerator and denominator of Eq. (\ref{Teff-}) both go to zero with their ratio tending
to a constant number. On the contrary, $T_{eff-}$ is a decreasing function of the number
of extra dimensions $n$.

The effective temperature $T_{eff-}$ was found to exhibit some unphysical properties, especially
in the case of charged de Sitter black holes where the aforementioned expression may take
on negative values or exhibit infinite jumps at the critical point. For this reason, in
\cite{Mann_review} (see also \cite{Shankar}) a new expression for the effective
temperature of a Schwarzschild-de Sitter spacetime was proposed, namely the following  
\beq
T_{eff+}=\left(\frac{1}{T_c}+\frac{1}{T_0}\right)^{-1}=\frac{T_0 T_c}{T_0 + T_c}\,.
\label{Teff+}
\eeq
The above proposal was characterised as an `ad hoc' one, that would follow from an
analysis similar to that leading to $T_{eff-}$ in which the entropy of the system would be
the difference of the entropies of the two horizons, i.e. $S=S_c-S_h$, instead of their sum. 
In the higher-dimensional case, the aforementioned alternative effective temperature has
the explicit form
\beq
T_{eff+}=\frac{1}{4\pi}\,\frac{(n+1)^2 -(n+1)(n+3) \tilde \Lambda (r_h^2+r_c^2)
+(n+3)^2 \tilde \Lambda^2 r_h^2 r_c^2}{(r_h -r_c)\,[(n+1) +(n+3) \tilde \Lambda r_h r_c]}\,.
\label{Teff+expl}
\eeq
In the limit $r_h \rightarrow 0$, $T_{eff+}$ reduces again to $T_c$. When $\Lambda \rightarrow
0$, it also exhibits the same behaviour as $T_{eff-}$ by going to zero. However, near the
critical point, $T_{eff+}$ has a distinct behaviour as it vanishes instead of taking a constant
value. This is in accordance with Eq. (\ref{Teff+}) where the numerator clearly approaches
zero faster than the denominator. It is perhaps the vanishing of $T_{eff+}$ near the
critical point that helps to avoid the infinite jumps and makes this alternative effective
temperature more physically acceptable. The complete behaviour of $T_{eff+}$ in terms
of the cosmological constant is  depicted in Fig. \ref{Temp_L_n1}; its decreasing behaviour
in terms of $n$ is also shown in Fig. \ref{Temp_n_L08}. 

Inspired by the above analysis, here we propose a third, alternative form for the effective
temperature of a Schwarzschild-de Sitter spacetime. Its functional form is the following
\beq
T_{effBH}=\left(\frac{1}{T_c}-\frac{1}{T_{BH}}\right)^{-1}=\frac{T_{BH} T_c}{T_{BH} - T_c}\,,
\label{TeffBH}
\eeq
and it matches the one of $T_{eff-}$, but with the normalised black-hole temperature
$T_{BH}$ in the place of the bare one $T_0$. Our proposal may be considered as an equally
`ad hoc' one compared to that of (\ref{Teff+}); however, $T_{effBH}$ would follow from exactly
the same analysis that gave rise to $T_{eff-}$ (with $S=S_h+S_c$) with the only difference 
being the consideration that the `correct' black-hole temperature, due to the absence of
asymptotic flatness, is $T_{BH}$ instead of $T_0$. Its explicit form in a spacetime with
$D=4+n$ dimensions is
\beq
T_{effBH}=-\frac{1}{4\pi}\,\frac{(n+1)^2 -(n+1)(n+3) \tilde \Lambda (r_h^2+r_c^2)
+(n+3)^2 \tilde \Lambda^2 r_h^2 r_c^2}{(r_h \sqrt{h(r_0)} +r_c)\,
[(n+1) -(n+3) \tilde \Lambda r_h r_c]}\,.
\label{TeffBH-expl}
\eeq
The above definition shares many characteristics with
the effective temperature $T_{eff-}$: it also reduces to $T_c$ when $r_h \rightarrow 0$
and it vanishes in the limit $\Lambda \rightarrow 0$. But it also exhibits the same attractive
behaviour near the critical point as $T_{eff+}$ by going to zero; this is due to the fact that,
as we approach the critical point, $T_{BH}$ in Eq. (\ref{TeffBH}) is a constant while $T_c$ vanishes.
The complete profile of $T_{effBH}$ as a function of the cosmological constant is depicted in
Fig. \ref{Temp_L_n1}, while its similar behaviour in terms of $n$, compared to the other effective 
temperatures, is shown in Fig. \ref{Temp_n_L08}. Observing Fig. \ref{Temp_L_n1}, it is interesting
to note that $T_{effBH}$ matches $T_{eff-}$ over an extended low $\Lambda$-regime, and then
coincides with $T_0$ in the high $\Lambda$-regime\footnote{One may wonder whether an alternative
effective temperature could be defined along the lines of Eq. (\ref{TeffBH}) but with a normalised
temperature for the cosmological horizon too, i.e. $T_{cBH}=T_c/\sqrt{h(r_0)}$. As one may see,
such a temperature would have a similar behaviour to $T_{eff-}$ in the small $\Lambda$-regime
but would have an ill-defined behaviour near the critical point where it diverges.}.


\section{Hawking Radiation for Minimally-Coupled Scalar Fields} 

In the previous section, we examined in detail the characteristics of two temperatures
for the Schwarzschild-de Sitter black hole, the bare $T_0$ and the normalised one
$T_{BH}$, as well as three effective temperatures for the Schwarzschild-de Sitter spacetime,
$T_{eff-}$, $T_{eff+}$ and $T_{effBH}$, to which the SdS black hole belongs. In this section,
we proceed to derive and compare the radiation spectra for scalar fields emitted by the SdS
black hole, for each one of the aforementioned five temperatures. 

Our analysis will focus on the higher-dimensional case and will present radiation spectra 
for scalar fields emitted both on the brane and in the bulk. To this end, we need also
the greybody factor for brane and bulk scalar fields propagating in the SdS background.
These have been derived analytically, in the limit of small cosmological constant, in \cite{KPP1}
and numerically, for arbitrary values of $\Lambda$, in \cite{KPP3}. Since here we are interested
in deriving the form of the spectra for the complete range of $\Lambda$, we will use the
exact results derived in \cite{KPP3}. For the sake of completeness, we will briefly review the
method for calculating the scalar greybody factors in a SdS spacetime - for more information,
interested readers may look in \cite{KPP3}.

We will start from the emission of scalar fields on the brane. The equation of motion
of a free, massless scalar field minimally-coupled to gravity and propagating in the brane
background (\ref{metric_brane}) has the form
\beq
\frac{1}{\sqrt{-g}}\,\partial_\mu\left(\sqrt{-g}\,g^{\mu\nu}\partial_\nu \Phi\right)=0\,. 
\label{field-eq-brane}
\eeq
If we assume a factorized ansatz for the field, i.e. 
$\Phi(t,r,\theta,\varphi)= e^{-i\omega t}\,R(r)\,Y(\theta,\varphi)$,
where $Y(\theta,\varphi)$ are the usual scalar spherical harmonics, we obtain a
radial equation for the function $R(r)$ of the form
\beq
\frac{1}{r^2}\,\frac{d \,}{dr} \biggl(hr^2\,\frac{d R}{dr}\,\biggr) +
\biggl[\frac{\omega^2}{h} -\frac{l(l+1)}{r^2}\biggr] R=0\,.
\label{radial_brane}
\eeq

As was shown in \cite{KPP3}, in the near-horizon regime, the above equation takes the form of
a hypergeometric equation. Its solution, when expanded in the limit $r \rightarrow r_h$
takes the form of an ingoing free wave, namely
\beq
R_{BH} \simeq A_1\,f^{\alpha_1} = A_1\,e^{-i(\omega r_h/A_h)\,\ln f}\,,
\label{BH-exp}
\eeq
where $A(r)=(n+1)-(n+3)\,\tilde \Lambda r^2$ and $A_h=A(r=r_h)$. Also,  $f$ is a new
radial variable defined through the relation
\beq
r \rightarrow f(r) = \frac{h(r)}{1- \tilde \Lambda r^2}\,.
\label{newco-f}
\eeq
For simplicity, we may appropriately choose the arbitrary constant $A_1$ so that 
\beq
R_{BH}(r_h)=1\,. \label{R_BH_num}
\eeq
The above expression serves as a boundary condition for the numerical integration of
Eq. (\ref{radial_brane}). The second boundary condition comes from the near-horizon
value of the first derivative of the radial function (\ref{BH-exp}) for which we obtain \cite{KPP3}
\beq 
\frac{dR_{BH}}{dr}\biggr|_{r_h} \simeq -\frac{i \omega}{h(r)}\,.
\label{dR_BH_num}
\eeq

Near the cosmological horizon, the radial equation (\ref{radial_brane}) takes again the form of
a hypergeometric differential equation whose general solution, in the limit $r \rightarrow r_c$
and $f \rightarrow 0$, is written as \cite{KPP1, KPP3}
\beq 
R_C \simeq  B_1\,e^{-i (\omega r_c/A_c)\ln f} + 
B_2\,e^{i (\omega r_c/A_c)\ln f} \,, \label{CO-exp}
\eeq
In the above, $A_c=A(r=r_c)$, and $B_{1,2}$ are the amplitudes of the ingoing and 
outgoing free waves. Then, the greybody factor, or equivalently the transmission probability,
for the scalar field is given by
\beq
|A|^2=1-\left|\frac{B_2}{B_1}\right|^2\,.
\label{greybody}
\end{equation}
The $B_{1,2}$ amplitudes are found by integrating numerically Eq. (\ref{radial_brane}),
starting close to the black-hole horizon, i.e. from $r=r_h+\epsilon$, where
$\epsilon=10^{-6}-10^{-4}$, and proceeding towards the cosmological horizon (again, for
more information on this, see \cite{KPP3}). The exact numerical analysis demonstrated
that for a minimally-coupled, massless scalar field propagating on the brane, the
greybody factor is enhanced over the whole energy regime as the cosmological constant
$\Lambda$ increases.

\begin{figure}[t]
  \begin{center}
\mbox{\includegraphics[width = 0.45 \textwidth] {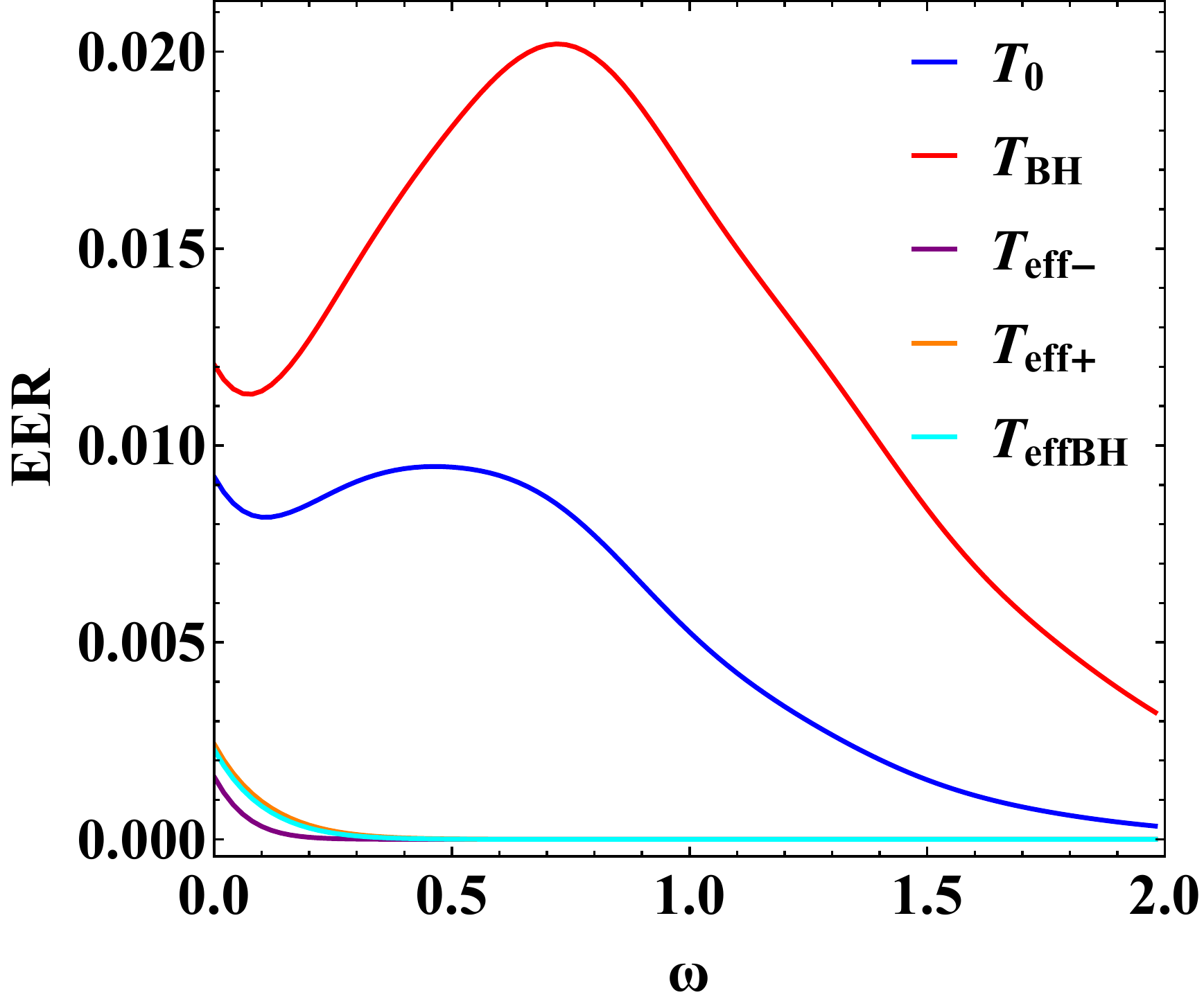}}
\hspace*{0.3cm} {\includegraphics[width = 0.45 \textwidth]
{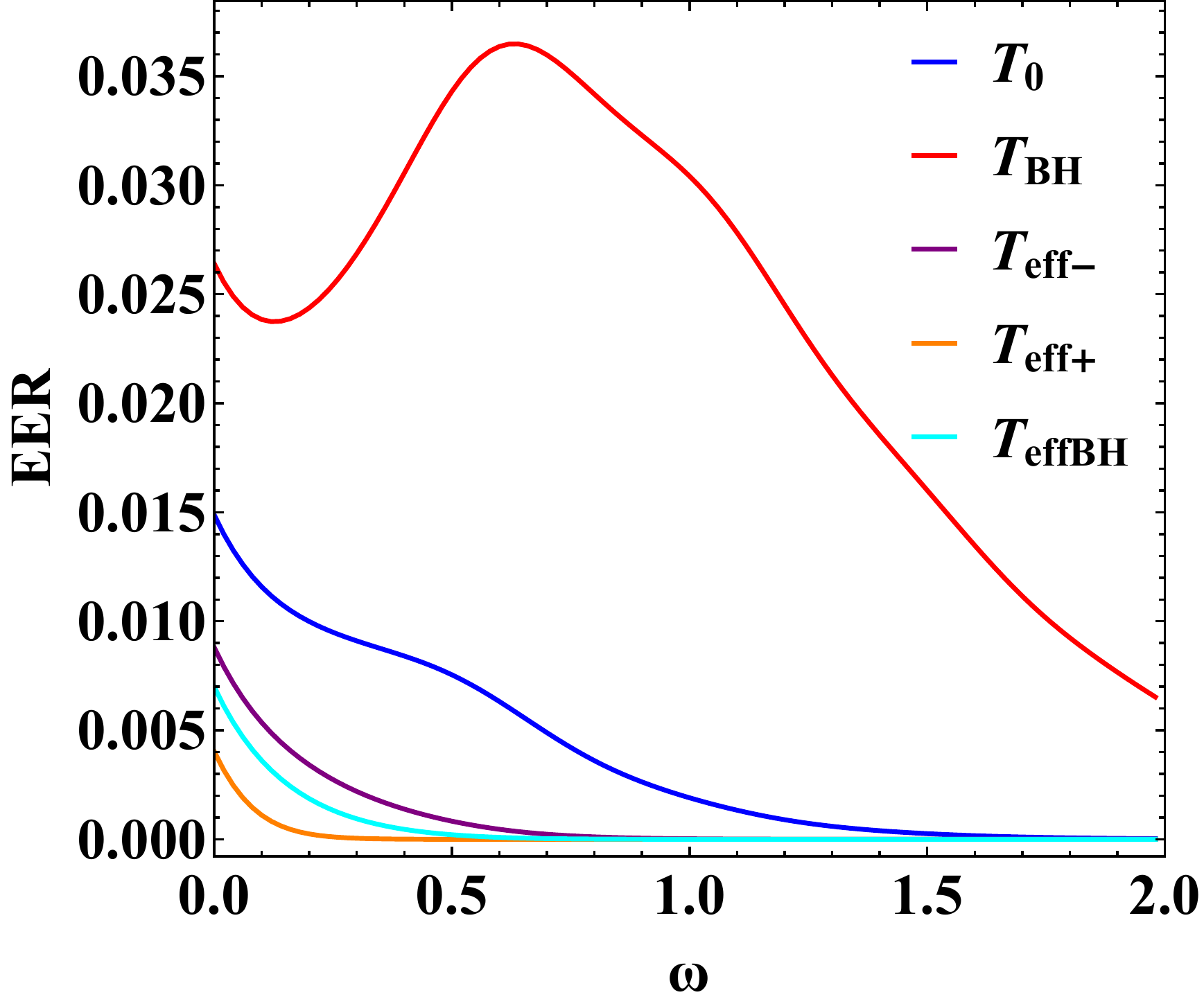}} \hspace*{0.3cm} {\includegraphics[width = 0.45 \textwidth]
{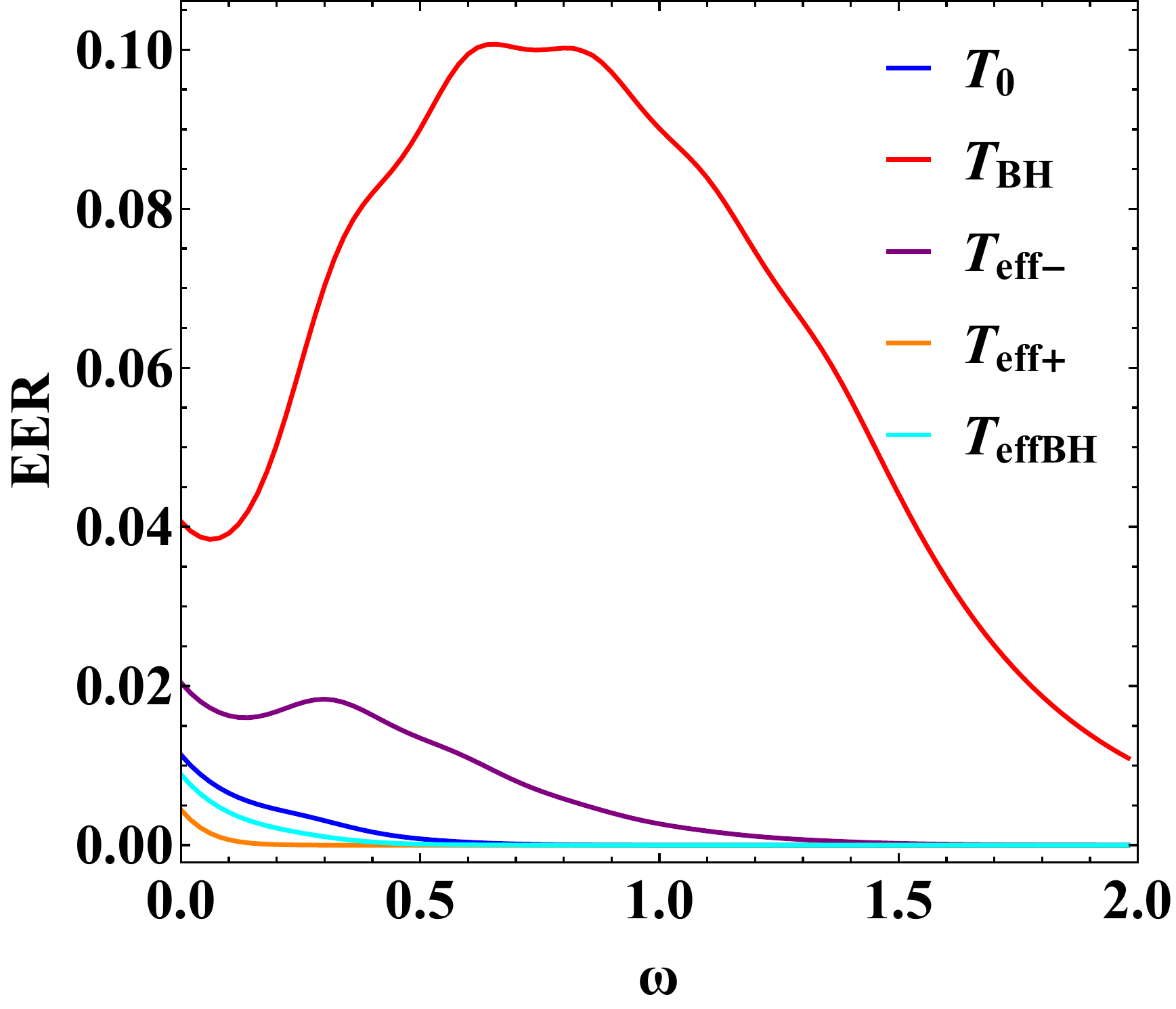}} \hspace*{0.3cm} {\includegraphics[width = 0.45 \textwidth]
{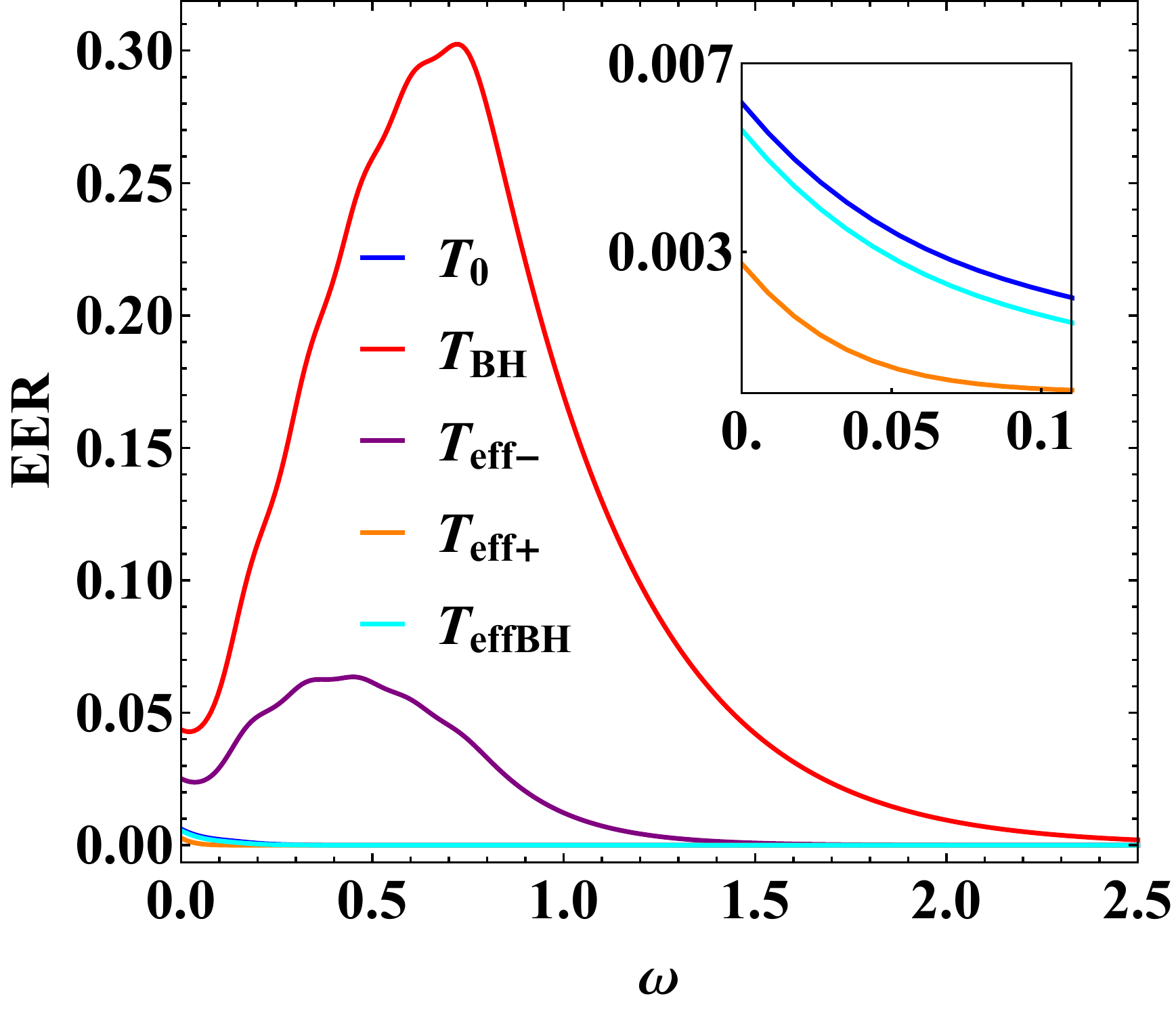}}
    \caption{Energy emission rates for scalar fields on the brane from a 6-dimensional
($n=2$) Schwarzschild-de Sitter black hole for different temperatures $T$, and for:
    \textbf{(a)} $\Lambda=0.8$, \textbf{(b)} $\Lambda=2$, \textbf{(c)} $\Lambda=4$,
and \textbf{(d)} $\Lambda=5$ (in units of $r_h^{-2}$).}
   \label{EER_brane_n2_L}
  \end{center}
\end{figure}

Having at our disposal the exact values of the greybody factor $|A|^2$, we may now proceed
to derive the differential energy emission rate for brane scalars. This is given by the expression
\cite{HK1, Kanti:2004, KGB}
\beq
\frac{d^2E}{dt\,d\omega}=\frac{1}{2\pi}\,\sum_l\,\frac{N_l\,|A|^2\,\omega}{\exp(\omega/T)-1}\,,
\label{diff-rate-brane}
\eeq
where $\omega$ is the energy of the emitted particle, and $N_l=2l+1$ the multiplicity of
states that, due to the spherical symmetry, have the same angular-momentum number \cite{Bander}.
Also, $T$ is the temperature of the black hole - this will be taken to be equal to
$T_0$, $T_{BH}$, $T_{eff-}$, $T_{eff+}$ and $T_{effBH}$, respectively, in order to derive the
corresponding radiation spectra. As was demonstrated in \cite{KPP3}, the dominant modes of
the scalar field are the ones with the lowest values of $l$ - in fact, all modes higher than
the $l = 7$ have negligible contributions to the total emission rate.

In Fig. \ref{EER_brane_n2_L}, we depict the differential energy emission rates for a higher-dimensional
Schwarzschild-de Sitter black hole for the case of $n=2$ and for four different values
of the bulk cosmological constant ($\Lambda=0.8, 2, 4, 5$). For the first, small value of
$\Lambda$, all the effective temperatures have an almost vanishing value, therefore the
corresponding spectra are significantly suppressed; it is the two black-hole temperatures,
$T_0$ and $T_{BH}$, that lead to significant emission rates, with the latter dominating over 
the former in accordance to the behaviour presented in Fig. \ref{Temp_L_n1}. As $\Lambda$
increases to the value of 2, the effective temperatures, and their corresponding spectra,
start becoming important; at the same time, the emission spectrum for the bare temperature
$T_0$ is suppressed whereas the one for the normalised $T_{BH}$ is enhanced.
For $\Lambda=4$ and $5$ finally, the radiation spectrum for $T_{BH}$ is further enhanced while
the one for $T_{eff-}$ has also become important - it is these two temperatures that
tend to a constant, non-vanishing value at the critical limit; on the contrary, all
three remaining temperatures, $T_0$, $T_{eff+}$ and $T_{effBH}$, tend to zero thus
causing a suppression to the corresponding spectra.

Let us also note that the traditional shape of the energy emission curves -- starting from
zero and reaching a maximum value before vanishing again -- is severely distorted. 
The presence of the cosmological constant leads to a non-vanishing asymptotic value 
of the greybody factor in the limit $\omega \rightarrow 0$ 
\cite{KGB, Harmark, Crispino, KPP1, KPP3}
given by
\beq
|A^2|=\frac{4 r_h^2r_c^2}{(r_c^2+r_h^2)^2}+ {\cal O}(\omega)\,.
\label{geom_brane}
\eeq
The above holds for the case of minimally-coupled, massless scalar fields propagating in
the brane background, and leads to a significant emission rate of extremely soft,
low-energetic particles -- this feature is evident in all plots of Fig. \ref{EER_brane_n2_L}.
In addition, when the temperature employed has a small value, like the effective temperatures
in the low and intermediate $\Lambda$-regime or $T_0$, $T_{eff+}$ and $T_{effBH}$ near
the critical limit, the emission curve never reaches a maximum at an energy larger
than zero; rather, it exhibits only the `tail', and monotonically decreases towards zero. 

\begin{figure}[t]
  \begin{center}
\mbox{\includegraphics[width = 0.45 \textwidth] {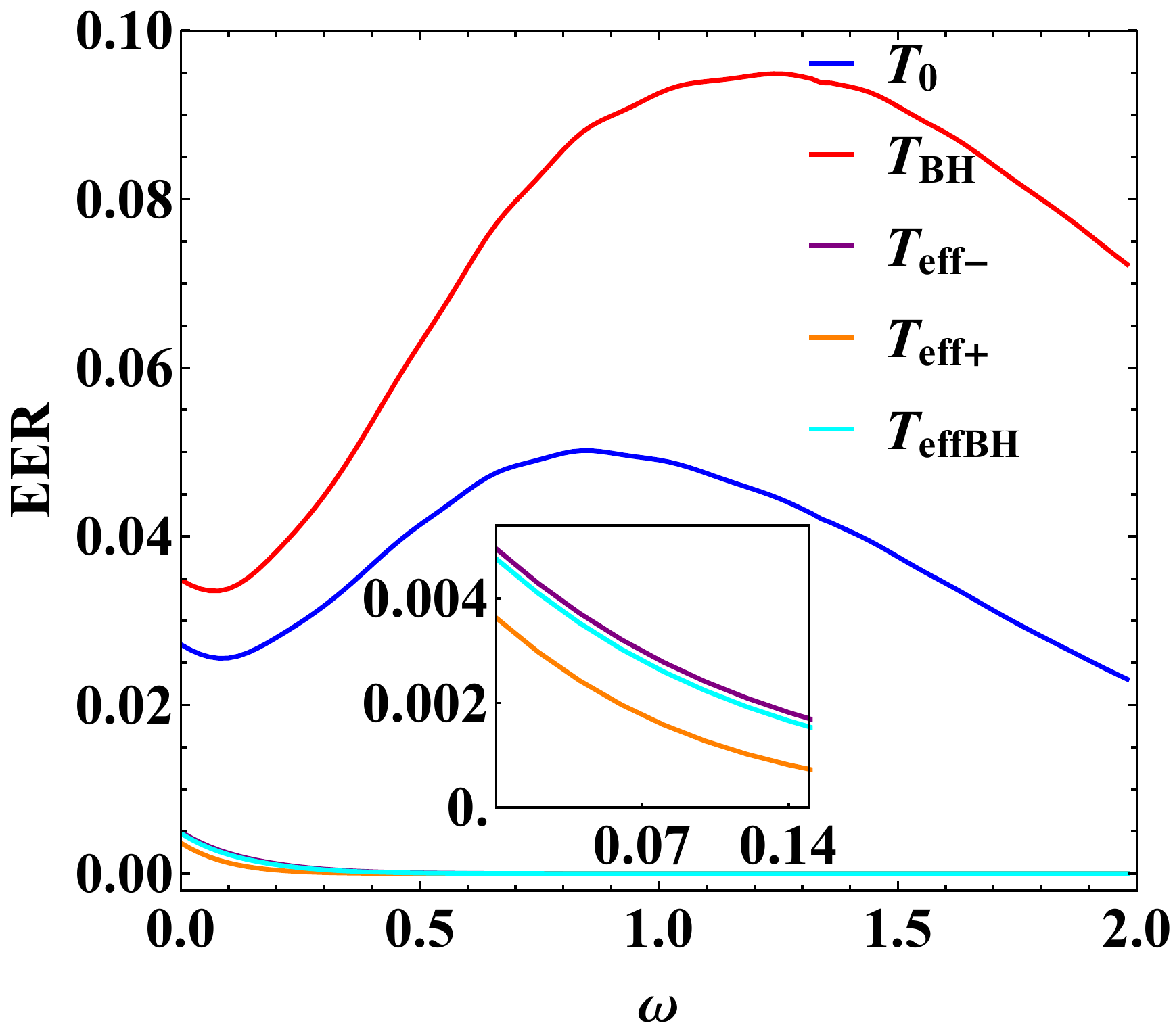}}
\hspace*{0.3cm} {\includegraphics[width = 0.45 \textwidth]
{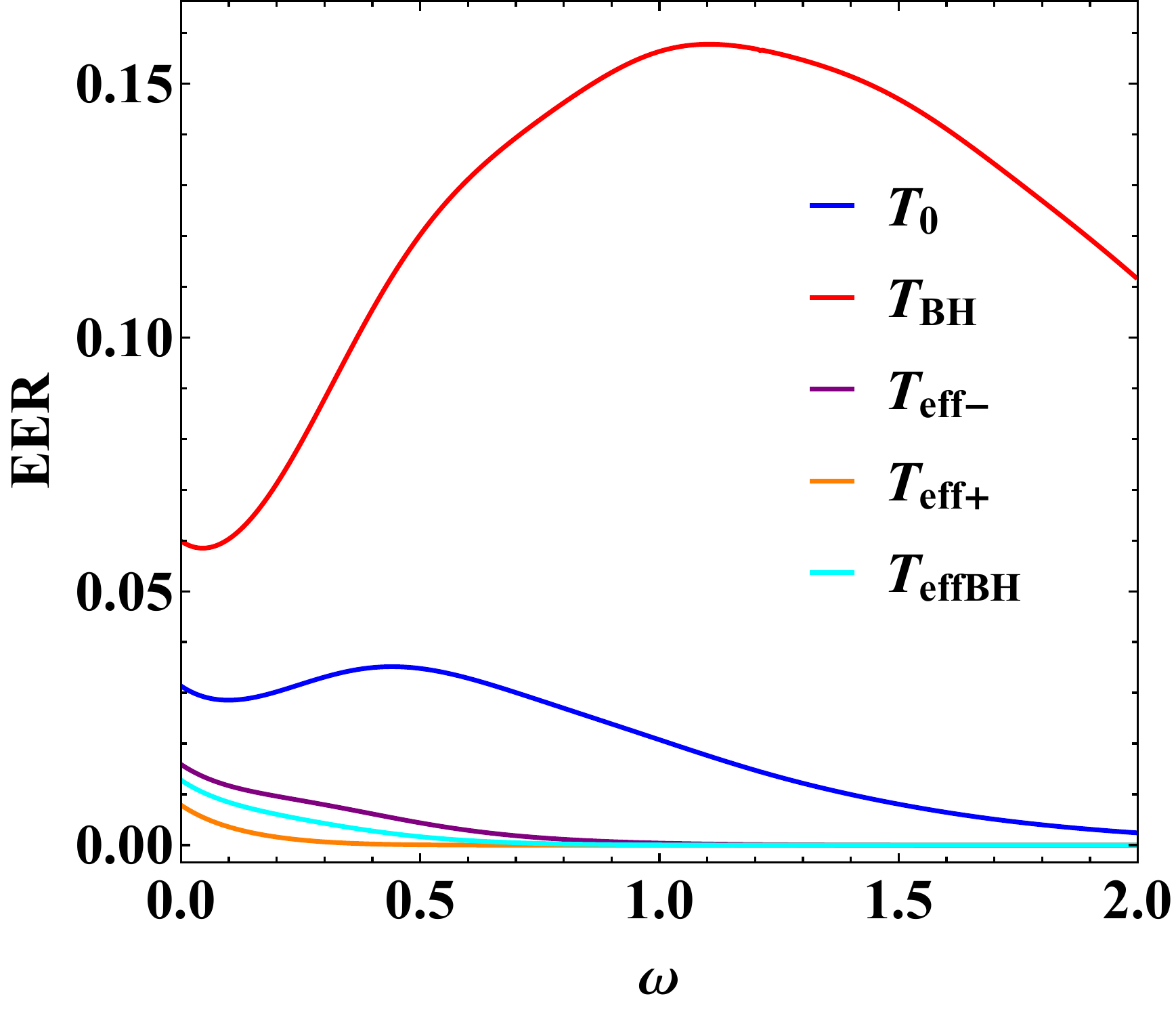}} \hspace*{0.3cm} {\includegraphics[width = 0.45 \textwidth]
{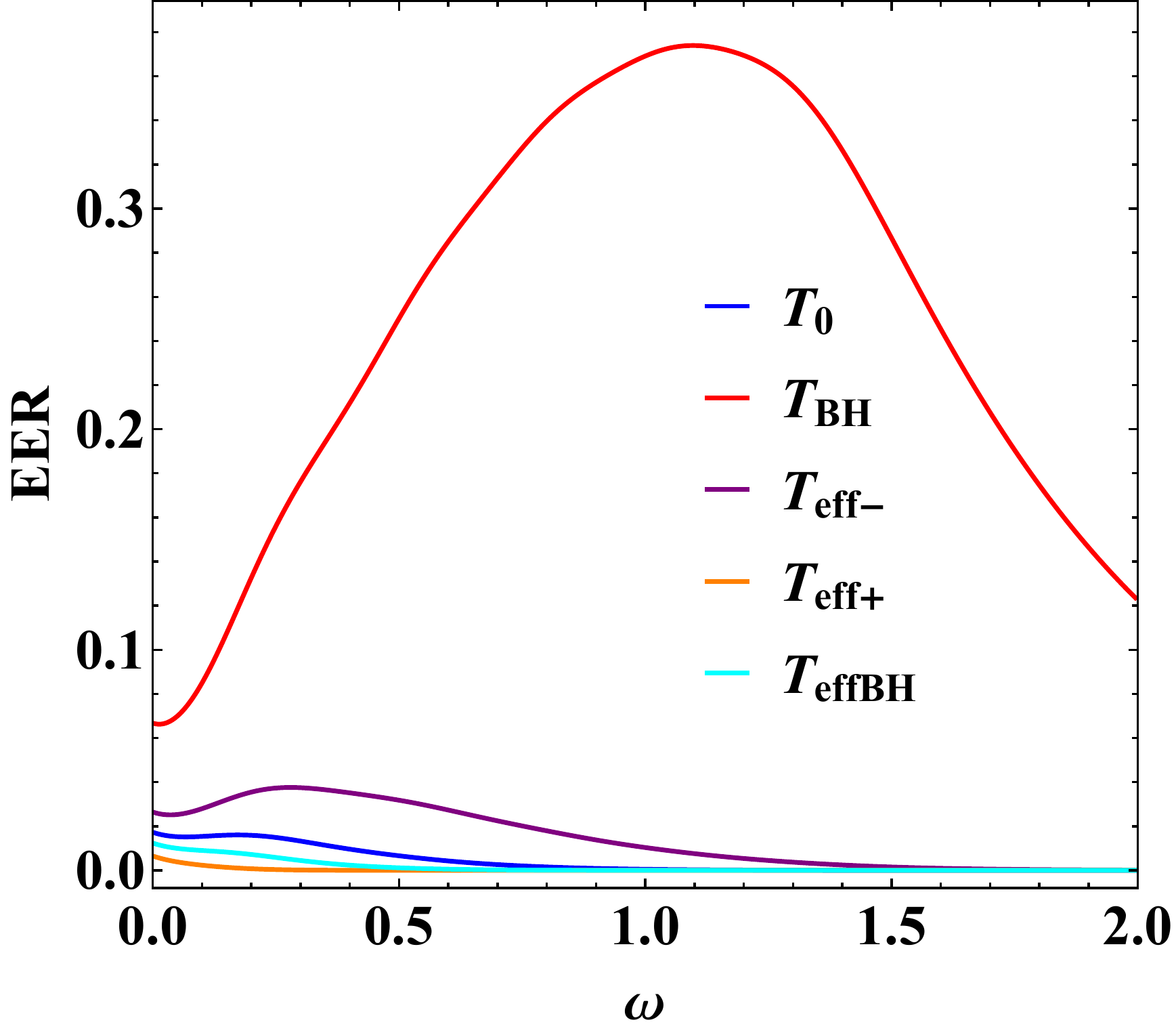}} \hspace*{0.3cm} {\includegraphics[width = 0.45 \textwidth]
{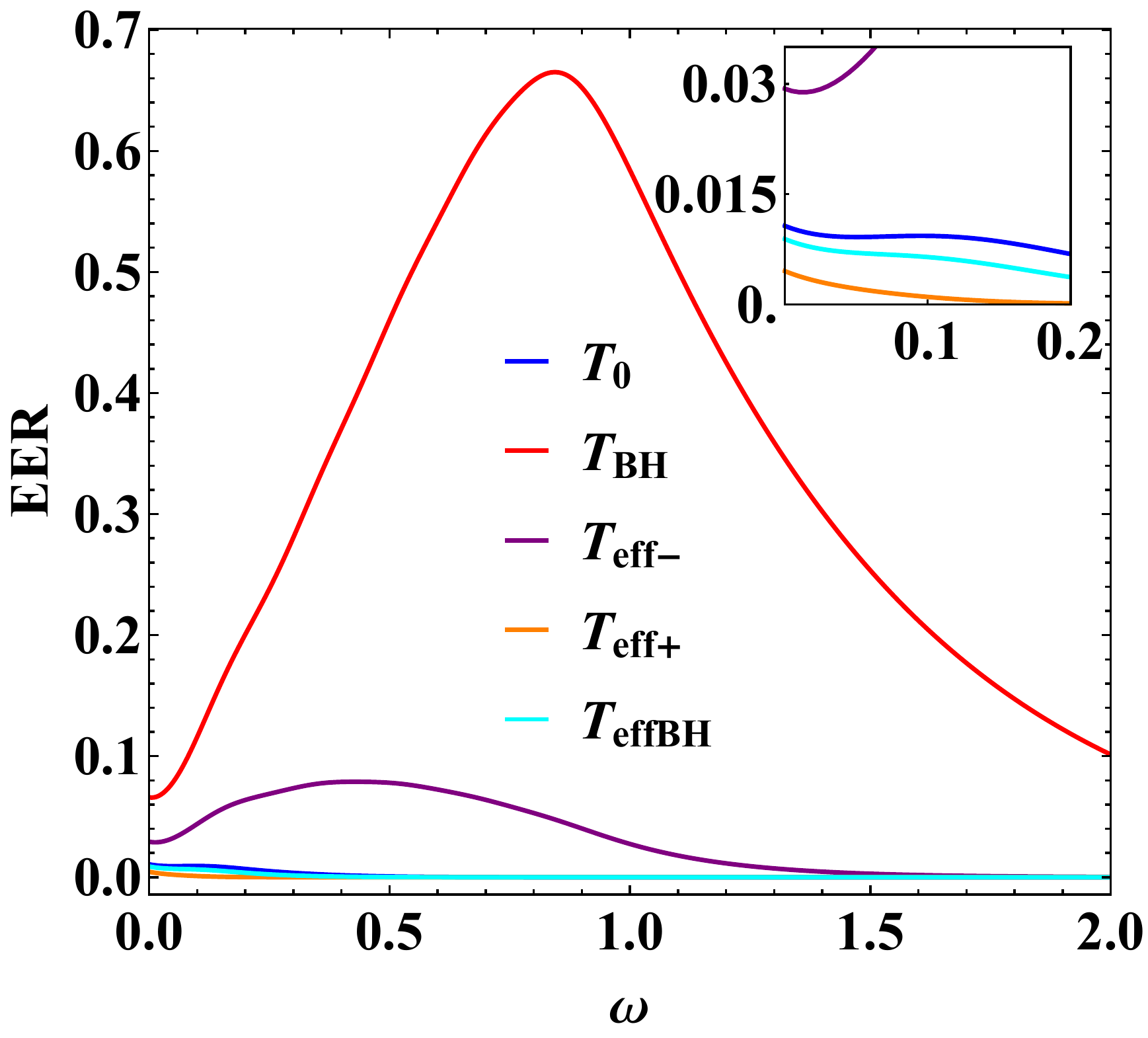}}
   \caption{Energy emission rates for scalar fields on the brane from a 9-dimensional
($n=5$) Schwarzschild-de Sitter black hole for different temperatures $T$, and for:
    \textbf{(a)} $\Lambda=4$, \textbf{(b)} $\Lambda=10$, \textbf{(c)} $\Lambda=16$,
 and \textbf{(d)} $\Lambda=18$ (in units of $r_h^{-2}$).}
   \label{EER_brane_n5_L}
  \end{center}
\end{figure}

The case of an even higher-dimensional Schwarzschild-de Sitter black hole with $n=5$ is
shown in Fig. \ref{EER_brane_n5_L}. A similar behaviour, to the one presented in the case
of $n=2$, is also observed here: for low values of $\Lambda$, the radiation spectra for
all effective temperatures are suppressed; as $\Lambda$
increases, they get moderately enhanced while for large values of $\Lambda$ only the
one for $T_{eff-}$ takes up significant values. The radiation spectrum for the bare
temperature $T_0$ starts
at its highest values for small $\Lambda$ and is constantly suppressed as the value
of the cosmological constant increases. The radiation spectrum for the normalised
black-hole temperature $T_{BH}$ is the one that dominates over the whole
$\Lambda$-regime -- even in the high $\Lambda$-regime, where $T_{BH}$ is
suppressed with $\Lambda$ according to Fig. \ref{Temp_L_n1}(b), the compensating
enhancement of the greybody factor \cite{KPP3} causes the overall increase of the
differential energy emission rate.

\begin{figure}[t]
  \begin{center}
\mbox{\includegraphics[width = 0.45 \textwidth] {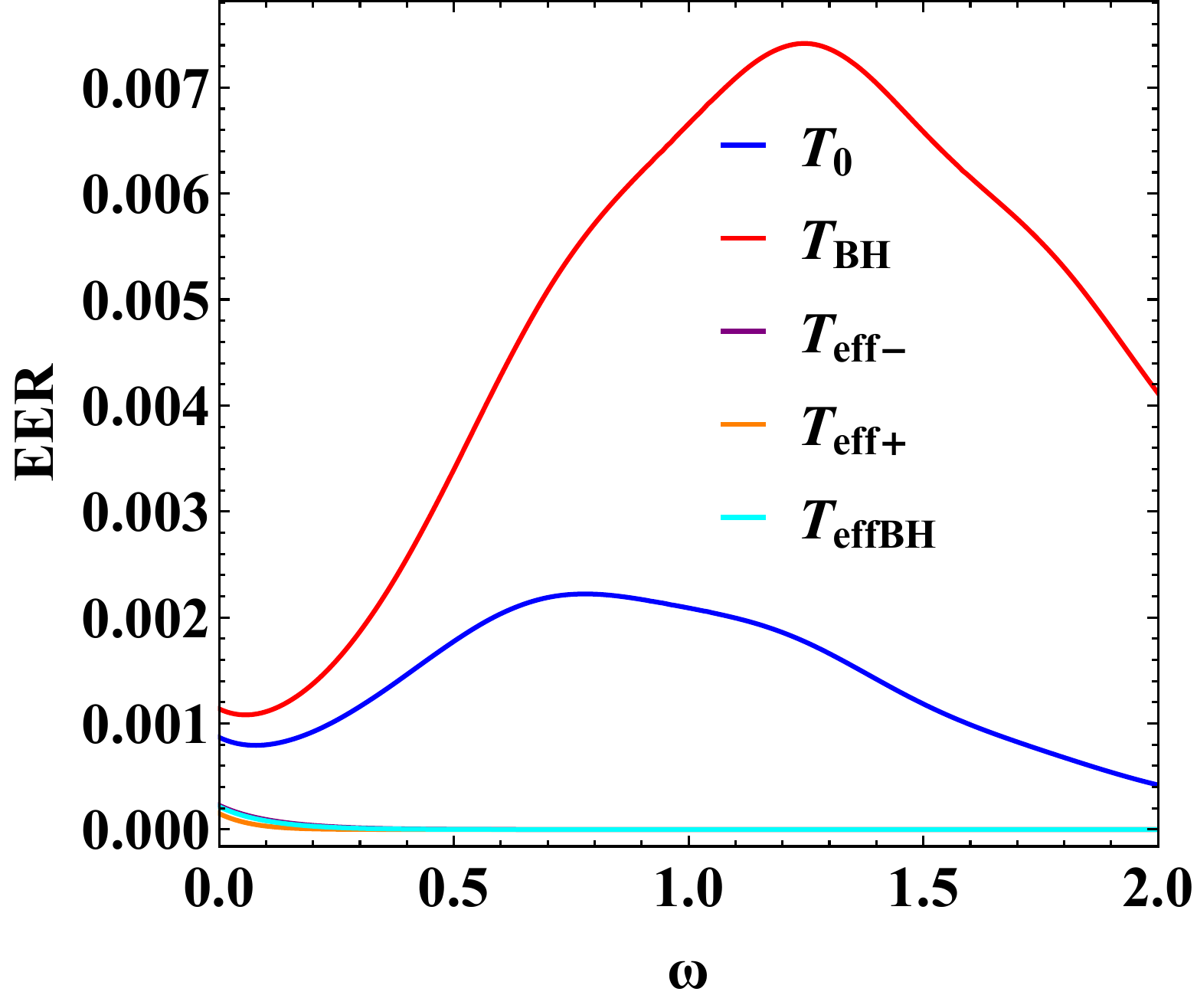}}
\hspace*{0.3cm} {\includegraphics[width = 0.45 \textwidth]
{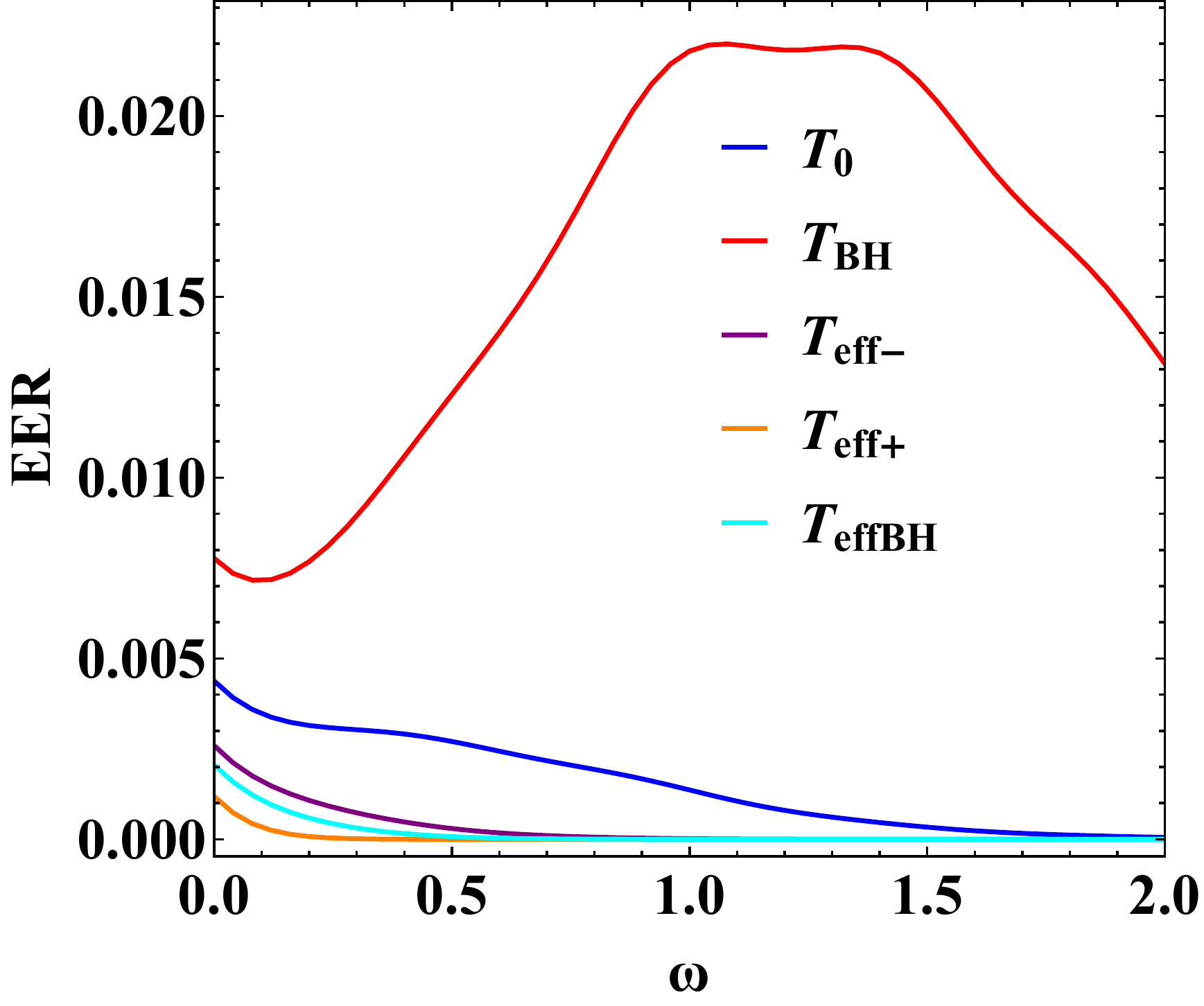}} \hspace*{0.3cm} {\includegraphics[width = 0.45 \textwidth]
{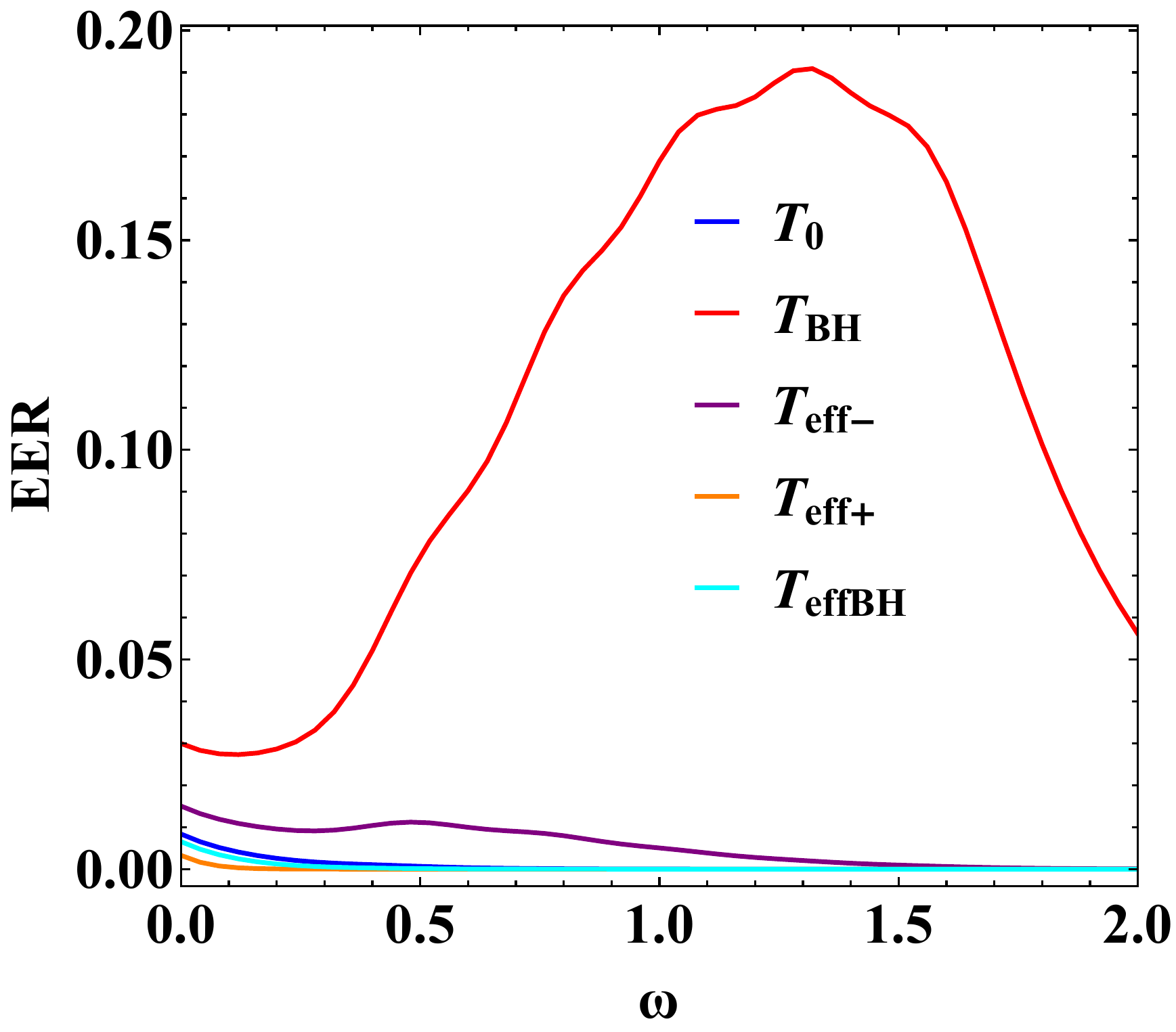}}  \hspace*{0.3cm} {\includegraphics[width = 0.45 \textwidth]
{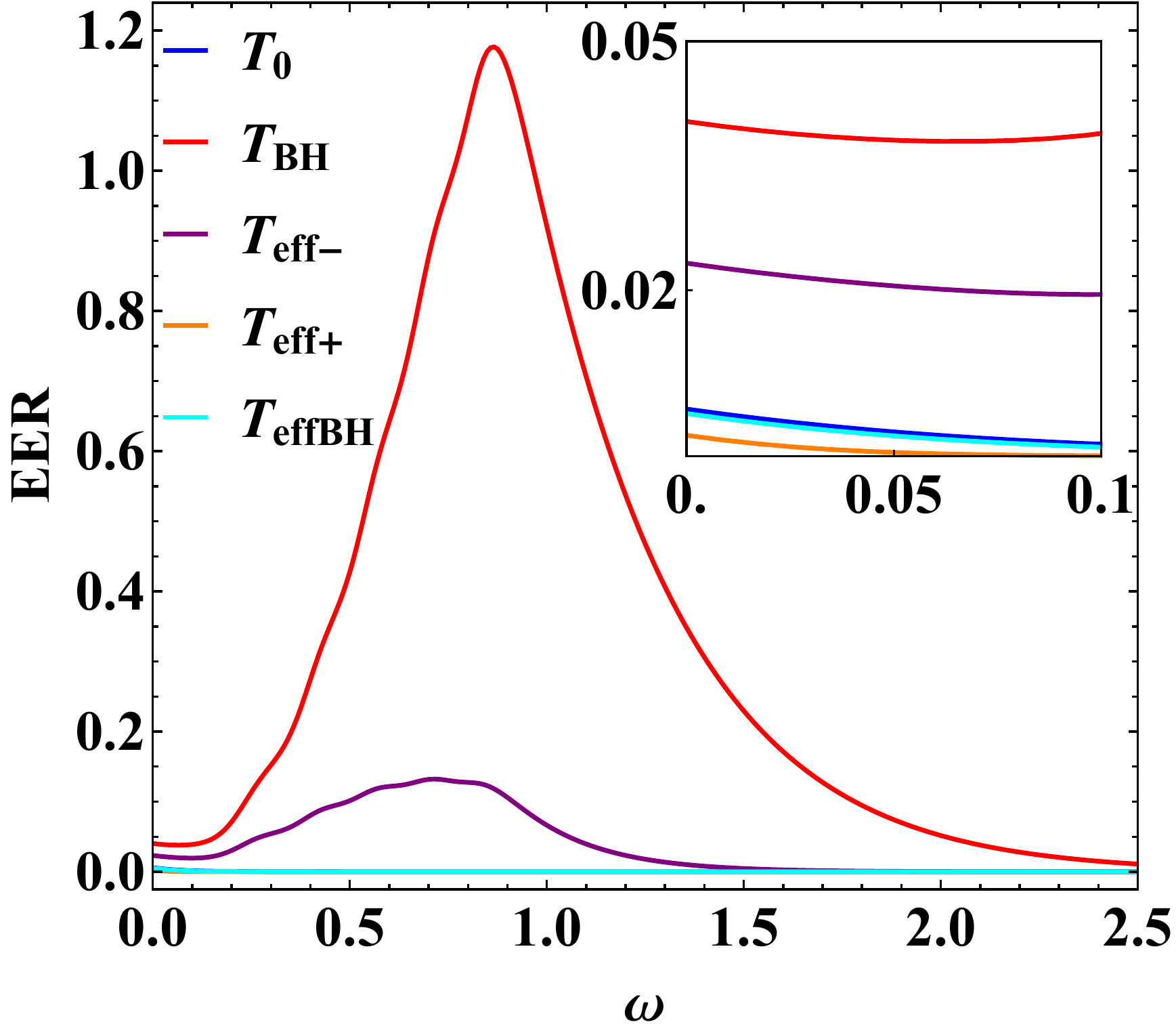}}
    \caption{Energy emission rates for scalar fields in the bulk from a 6-dimensional
($n=2$) Schwarzschild-de Sitter black hole for different temperatures $T$, and for:
    \textbf{(a)} $\Lambda=0.8$, \textbf{(b)} $\Lambda=2$, \textbf{(c)}
    $\Lambda=4$,  and \textbf{(d)} $\Lambda=5$ (in units of $r_h^{-2}$).}
   \label{Bulk_EER_n2_L}
  \end{center}
\end{figure}

Let us also study the emission of scalar fields from a higher-dimensional
Schwarz- schild-de Sitter black hole in the bulk. The equation of motion of
a free, massless field propagating in the bulk is also given by the covariant
equation (\ref{field-eq-brane}) but with the projected metric tensor $g_{\mu\nu}$
of Eq. (\ref{metric_brane}) being replaced by the higher-dimensional one $G_{MN}$
given in Eq. (\ref{bhmetric}). Assuming again a factorized form 
$\Phi(t,r,\theta_i,\varphi) = e^{-i\omega t}R(r)\,\tilde Y(\theta_i,\varphi)$,
where $\tilde Y(\theta_i,\varphi)$ are the hyperspherical harmonics \cite{Muller},
we obtain the following radial equation \cite{KPP1}
\beq
\frac{1}{r^{n+2}}\,\frac{d \,}{dr} \biggl(hr^{n+2}\,\frac{d R}{dr}\,\biggr) +
\biggl[\frac{\omega^2}{h} -\frac{l(l+n+1)}{r^2}\biggr] R=0\,.
\label{radial_bulk}
\eeq
The above differential equation may be again analytically solved for small
$\Lambda$ \cite{KPP1} but, for the purpose of comparing the radiation spectra
over the entire $\Lambda$ regime, we turn again to numerical integration.
This has been performed in \cite{KPP3} by following an analysis similar
to the one for brane scalar fields. The asymptotic solutions of
Eq. (\ref{radial_bulk}) near the black-hole and cosmological horizons
take similar forms to the brane ones, with their expanded forms 
(\ref{BH-exp}) and (\ref{CO-exp}) being identical. The same boundary
conditions (\ref{R_BH_num})-(\ref{dR_BH_num}) were used for the numerical
integration from the black-hole to the cosmological horizon. The exact value
of the greybody factor for bulk scalar fields, for arbitrary values of the
particle and spacetime parameters, was again derived via Eq. (\ref{greybody}),
and found to be an increasing function of the bulk cosmological constant.

\begin{figure}[t]
  \begin{center}
\mbox{\includegraphics[width = 0.45 \textwidth] {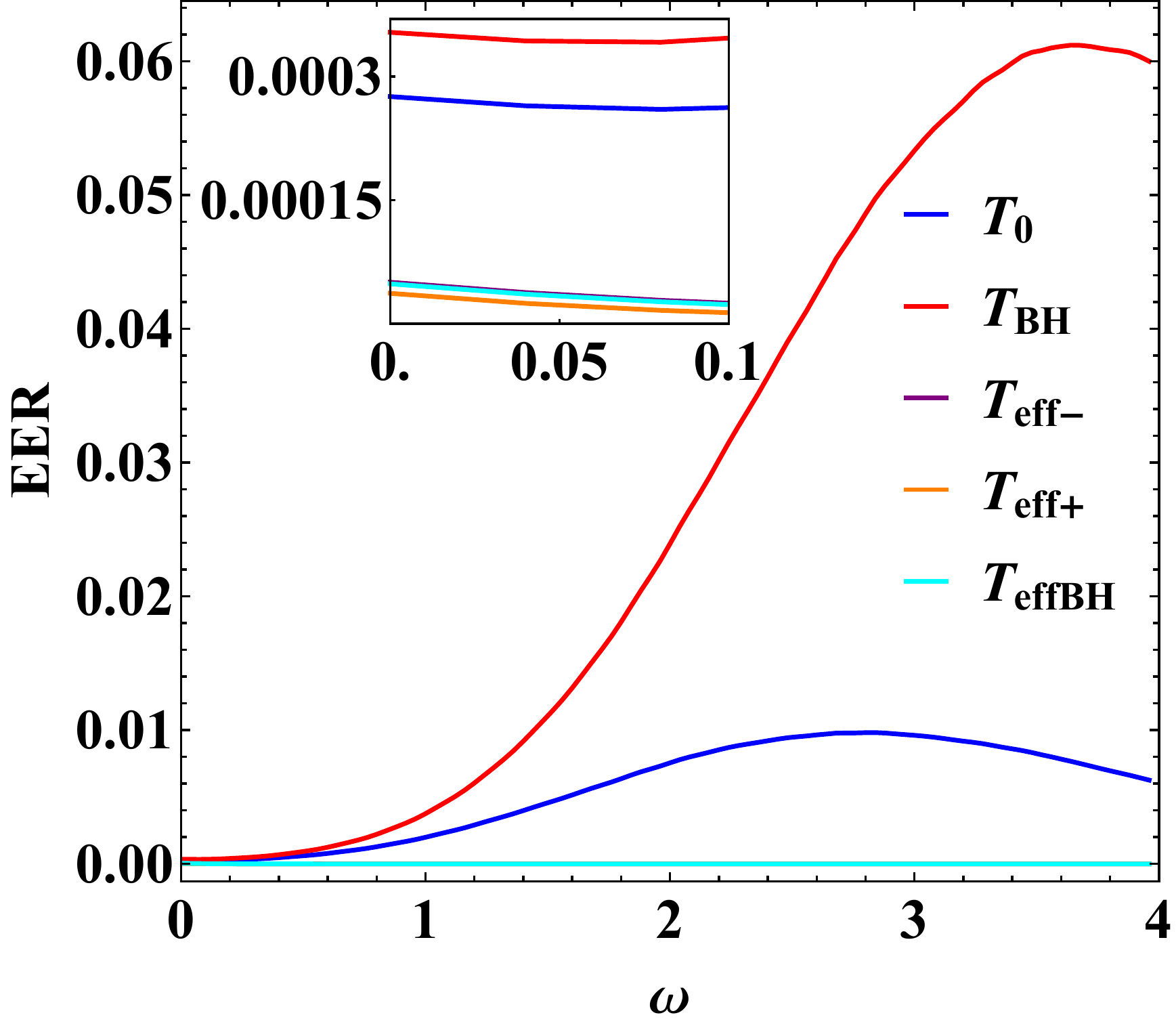}}
\hspace*{0.3cm} {\includegraphics[width = 0.45 \textwidth]
{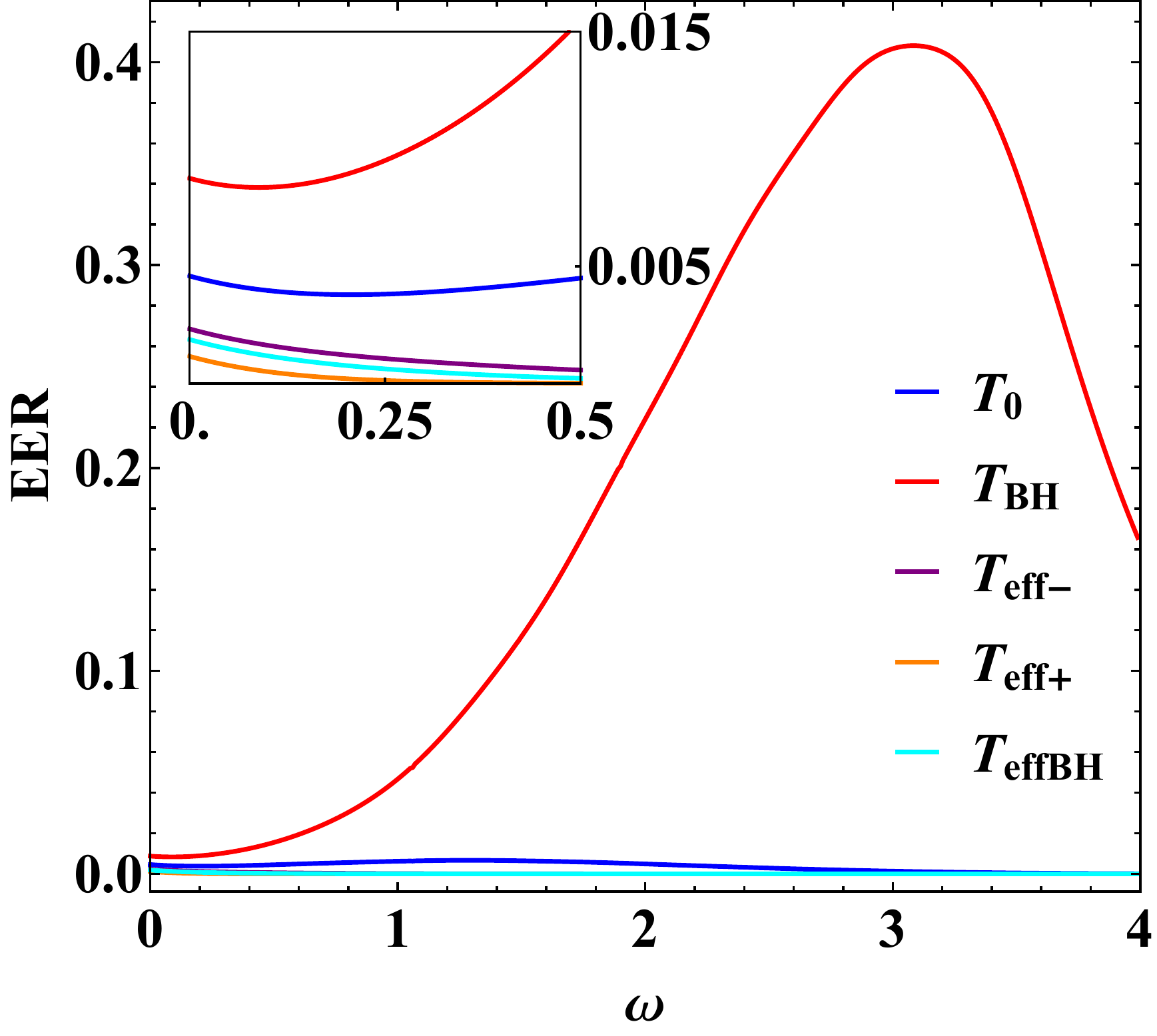}} \hspace*{0.3cm} {\includegraphics[width = 0.45 \textwidth]
{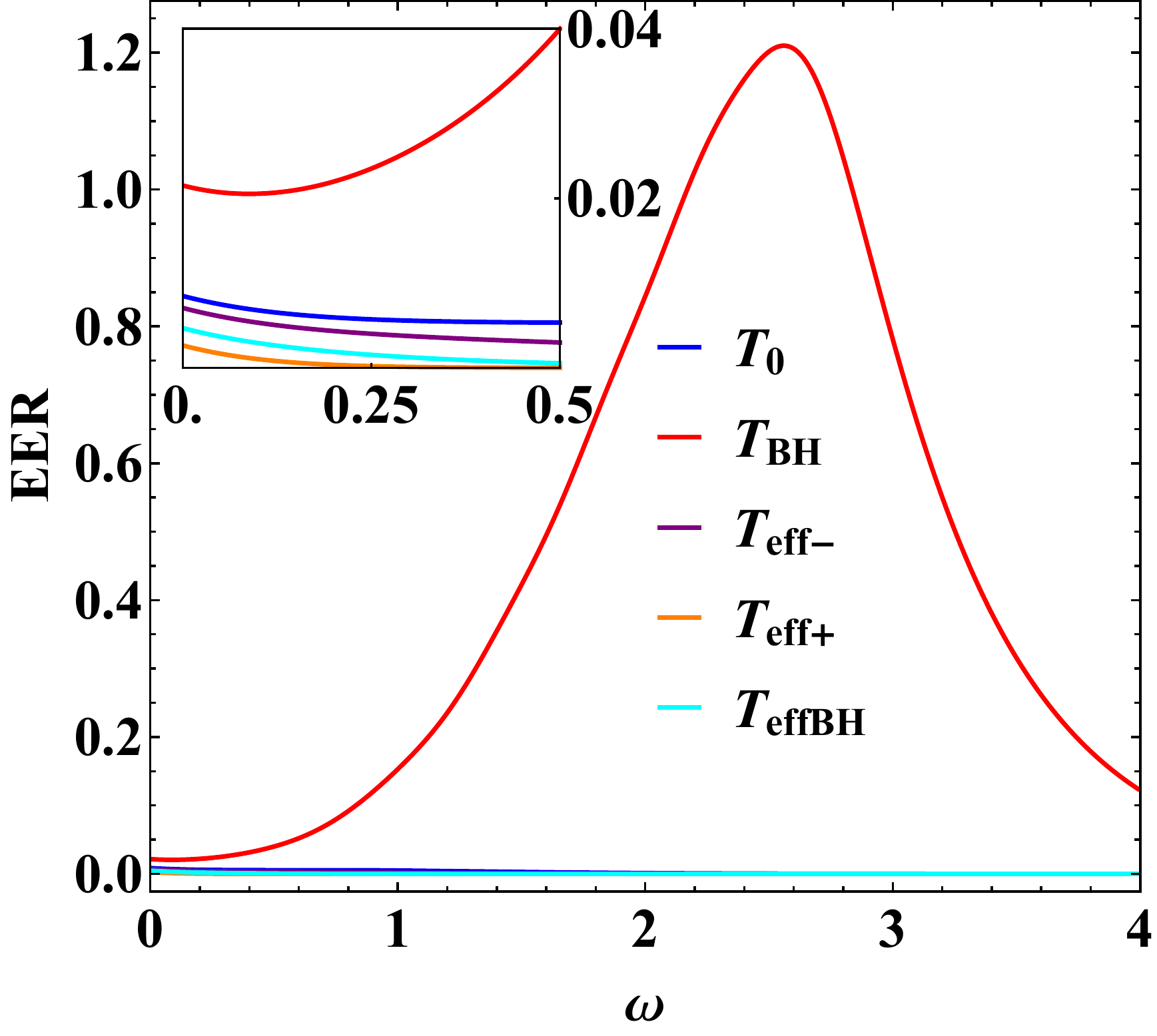}} \hspace*{0.3cm} {\includegraphics[width = 0.45 \textwidth]
{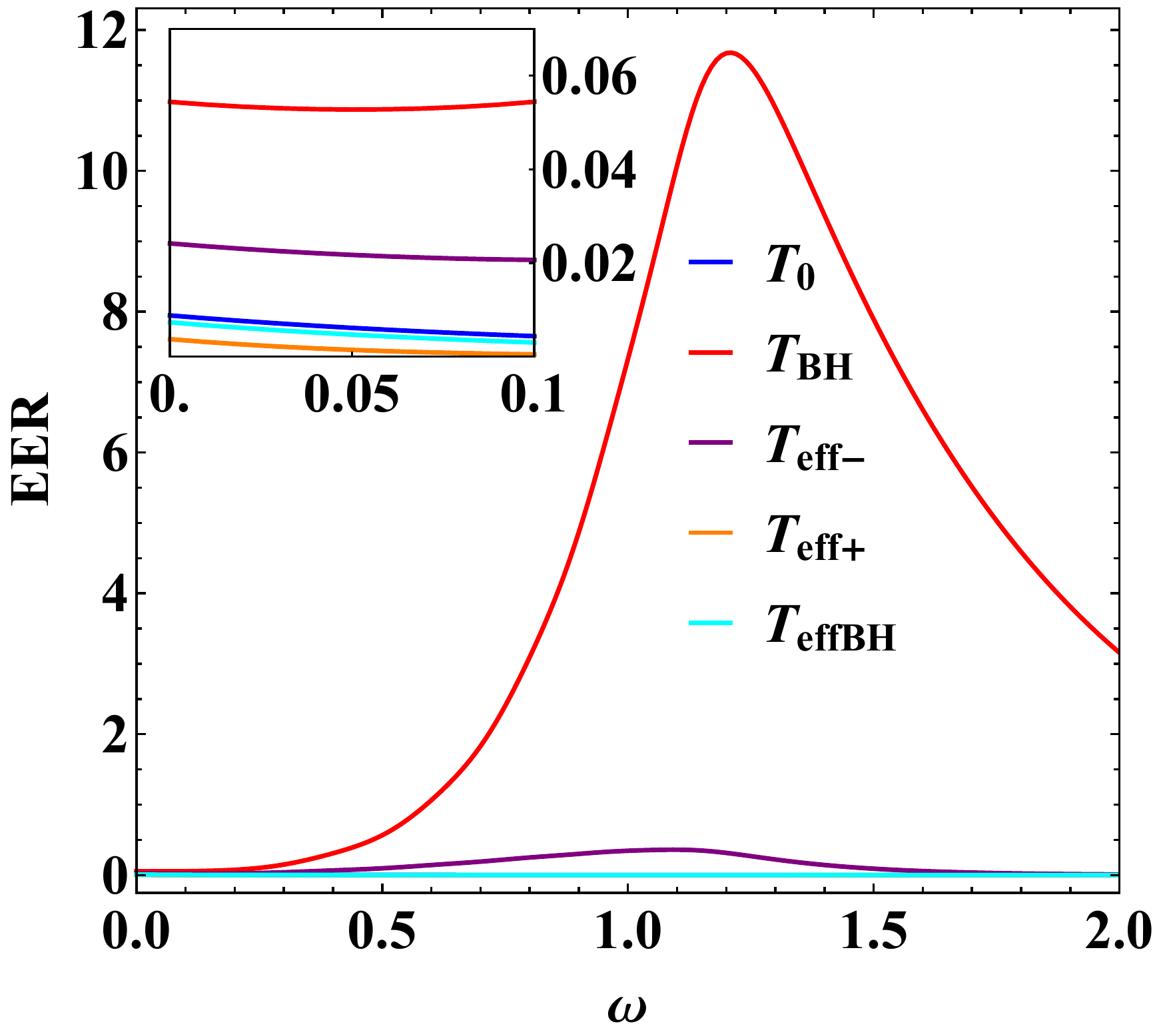}}
    \caption{Energy emission rates for scalar fields in the bulk from a 9-dimensional
($n=5$) Schwarzschild-de Sitter black hole for different temperatures $T$, and for:
    \textbf{(a)} $\Lambda=4$, \textbf{(b)} $\Lambda=10$, \textbf{(c)}
    $\Lambda=13$ and \textbf{(d)} $\Lambda=18$ (in units of $r_h^{-2}$).}
   \label{Bulk_EER_n5_L}
  \end{center}
\end{figure}

In Fig. \ref{Bulk_EER_n2_L}, we display the differential energy emission rates
for bulk scalar fields emitted by a 6-dimensional ($n=2$) SdS black hole, and
for four different values of the cosmological constant. Similarly to the
behaviour observed in the case of brane emission, the radiation spectrum for
the normalised temperature $T_{BH}$ is the one that dominates and gets
enhanced as $\Lambda$ increases, under the combined effect of the temperature
and greybody profiles. The spectrum for the bare temperature $T_0$, starting
from significant values for low $\Lambda$, is again monotonically suppressed
as $\Lambda$ increases approaching its maximum critical value. The spectra
for all effective temperatures start from extremely low values and only the
one for $T_{eff-}$ manages to
reach non-negligible values -- this takes place only near the critical
limit where $T_{eff-}$ acquires a constant value. If we allow for a larger
value of the number of extra dimensions, i.e. $n=5$, the general behaviour of
the emission curves remains the same, as can be seen from the plots in
Fig. \ref{Bulk_EER_n5_L}, drawn for four different values of the cosmological constant.
Here, the additional suppression of all effective temperatures with $n$
keeps even more the corresponding radiation spectra at low values.  

Also in the bulk, all emission curves tend to a non-vanishing value in the limit
$\omega \rightarrow 0$. This is again due to the non-zero asymptotic value of the
greybody factor at the very low-energy regime. However, this value for bulk emission
is \cite{KGB} 
\beq
|A^2|=
\frac{4 (r_hr_c)^{(n+2)}}{(r_c^{n+2}+r_h^{n+2})^2}+ {\cal O}(\omega)\,.
\label{geom_bulk}
\eeq
The above expression is suppressed with the number of extra dimensions $n$ and this is
the reason why this feature is more difficult to discern in the 9-dimensional emission
curves of Fig. \ref{Bulk_EER_n5_L} compared to the 6-dimensional ones of 
Fig. \ref{Bulk_EER_n2_L} -- it is nevertheless visible in the zoom-in plots 
that have been added in Fig. \ref{Bulk_EER_n5_L}. 

The numerical analysis performed in the context of the present work serves not only
as a comparison of the radiation emission curves when different expressions for the
temperature of the SdS spacetime are used, but also as an extension to the previous
results obtained in \cite{KPP3} where the normalised temperature $T_{BH}$ was employed.
There, exact results for the radiation spectra were produced but the range of values of
the cosmological constant was much more restricted, i.e. $\Lambda \in [0.01, 0.3]$, 
therefore, the regime of large values of $\Lambda$, including the critical limit, was
never studied. Here, we have performed a thorough analysis of the $\Lambda$ regime
for all different temperatures, and thus we have the complete picture of how the
corresponding radiation spectra behave as a function of the value of the cosmological
constant. 

Overall, after having performed both the brane and the bulk analysis, we may conclude that
it is the black-hole temperatures $T_0$ and $T_{BH}$ that lead to Hawking radiation emission
curves with the typical shape, i.e. start from a low value at the low-energy regime, rise to a
maximum height and then slowly die out at the high-energy regime. In fact, even the
$T_0$ spectrum loses this typical shape as $\Lambda$ increases. Of the effective
temperatures, only $T_{eff-}$ manages to mimic this behaviour, and does so only close to
the critical limit. 

If we focus on the most typical radiation spectra, i.e. the ones derived for the normalised
temperature $T_{BH}$, we could comment on some additional features that emerge from
the more thorough
study, in terms of the $\Lambda$-regime, performed in the present work. Our current
results have confirmed the enhancement of the corresponding radiation spectra in terms
of both the number of extra dimensions $n$ and the value of the cosmological constant,
as found in \cite{KPP3}. As $\Lambda$ increases, the non-zero asymptotic value of each
curve in the limit $\omega \rightarrow 0$ is enhanced thus increasing the probability
of the emission of very low-energetic particles. In addition, for large values of $n$, 
as $\Lambda$ increases, all emission curves, for brane and bulk propagation alike, show a
significant shift of the peak of the curves towards the lower part of the spectrum. Therefore,
we may conclude that the presence of a cosmological constant gives a significant boost
to both low and intermediate-energy free, massless scalar particles and it does so more
effectively the larger the number of extra dimensions is. 

Finally, in \cite{KPP3} it was found that for $\Lambda$ in the regime 
$[0.01,0.3]$, the brane emission channel for free, massless scalar fields is always dominant 
compared to the bulk channel. Here, we observe that for larger values of $\Lambda$
the situation is radically changed: even for small values of $n$, i.e. $n=2$, the comparison
of the vertical axes of the plots of Figs. \ref{EER_brane_n2_L} and \ref{Bulk_EER_n2_L}
reveals that the bulk emission curve has surpassed, by a factor of two, the brane one,
for values of $\Lambda$ larger than 4. As the dimensionality of spacetime increases, the
bulk dominance becomes more important: for $n=5$, the comparison of the vertical
axes of the plots of  Figs. \ref{EER_brane_n5_L} and \ref{Bulk_EER_n5_L}, now tells us
that the bulk dominates over the brane for values of $\Lambda > 10$, i.e. for more
than half the allowed regime of values of the cosmological constant, by a factor that
ranges between 3 and 20.


\section{Hawking Radiation Spectra for Non Minimally-Coupled Scalar Fields}

In this section, we will consider the case of scalar particles propagating
either on the brane or in the bulk and having a non-minimal coupling to
gravity. This coupling is realised through a quadratic function $\xi \Phi^2$,
where $\xi$ is a constant, multiplying the appropriate scalar curvature
(with the value $\xi=0$ corresponding to the minimal coupling). 
The reason for studying such a theory is two-fold: first, the presence of
the non-minimal coupling acts as an effective mass term for the scalar field,
therefore the effect of the mass on the radiation spectra may thus be studied;
second, for large values of the coupling constant $\xi$, it was found that
the enhancement of the radiation spectra with the cosmological constant --
for the normalised temperature $T_{BH}$, that was also evident in the
results of the previous section -- changes to a suppression in the
low-energy and intermediate-energy regimes \cite{KPP3}. It would thus be
interesting to see what the effect of the non-minimal coupling would be on
the radiation spectra over for the complete $\Lambda$ regime and for different
temperatures.
 
For a scalar field propagating in the bulk, its higher-dimensional action
would read
\beq
S_\Phi=-\frac{1}{2}\,\int d^{4+n}x \,\sqrt{-G}\left[\xi \Phi^2 R_D
+\partial_M \Phi\,\partial^M \Phi \right]\,,
\label{action-scalar-bulk}
\eeq
where $G_{MN}$ is again the higher-dimensional metric tensor defined in
Eq. (\ref{bhmetric}), and $R_D$ the corresponding curvature given by the
expression
\beq
R_D=\frac{2\,(n+4)}{n+2}\,\kappa^2_D \Lambda\,,
\label{RD}
\eeq
in terms of the bulk cosmological constant. The equation of motion of the bulk scalar field
now reads
\beq
\frac{1}{\sqrt{-G}}\,\partial_M\left(\sqrt{-G}\,G^{MN}\partial_N \Phi\right)
=\xi R_D\,\Phi\,,
\label{field-eq-bulk-xi}
\eeq
or, more explicitly,
\beq
\frac{1}{r^{n+2}}\,\frac{d \,}{dr} \biggl(hr^{n+2}\,\frac{d R}{dr}\,\biggr) +
\biggl[\frac{\omega^2}{h} -\frac{l(l+n+1)}{r^2}-\xi R_D\biggr] R=0\,.
\label{radial_bulk-xi}
\eeq
In the above we have decoupled the radial part of the equation by considering
the same factorized ansatz, namely
$\Phi(t,r,\theta_i,\varphi) = e^{-i\omega t}R(r)\,\tilde Y(\theta_i,\varphi)$,
as in the previous section. 

The action functional for a scalar field propagating on the brane background and having
also a quadratic, non-minimal coupling to the scalar curvature, will have a form similar 
to Eq. (\ref{action-scalar-bulk}). However, now the metric tensor $G_{MN}$ will be replaced
by the projected-on-the-brane one $g_{\mu\nu}$ given in Eq. (\ref{metric_brane}), and the
higher-dimensional Ricci scalar $R_D$ by the four-dimensional one $R_4$ that is found
to be \cite{KPP1}
\beq
R_4=\frac{24 \kappa_D^2 \Lambda}{(n+2) (n+3)} + \frac{n(n-1)\mu}{r^{n+3}}\,.
\label{R4}
\eeq
The equation for the radial part of the brane-localised, non-minimally coupled scalar field 
then follows from Eq. (\ref{radial_bulk-xi}) by setting $n=0$ and changing $R_D$ with
$R_4$, and reads
\beq
\frac{1}{r^2}\,\frac{d \,}{dr} \biggl(hr^2\,\frac{d R}{dr}\,\biggr) +
\biggl[\frac{\omega^2}{h} -\frac{l(l+1)}{r^2}-\xi R_4\biggr] R=0\,.
\label{radial_brane-xi}
\eeq

Both equations (\ref{radial_bulk-xi}) and (\ref{radial_brane-xi}) were solved analytically
in \cite{KPP1} and numerically in \cite{KPP3}. As it is clear from both equations, the 
non-minimal coupling term acts as an effective mass term, therefore any increase in
the coupling function $\xi$ causes a suppression to the radiation spectra, in accordance
to previous studies of massive scalar fields \cite{Page, Jung, Sampaio, KNP1}. In
addition, in \cite{KPP3},
it was found that as $\xi$ exceeds the value of approximately 0.3, any increase in the
value of the cosmological constant causes a suppression in the low and intermediate
part of the spectrum. 

In the light of the above, here we will consider a value for the non-minimal coupling constant
well beyond that critical value, namely we will choose $\xi=1$. We will also study the 
complete $\Lambda$-regime and compute the radiation spectra for all five temperatures,
$T_{0}$, $T_{BH}$, $T_{eff-}$, $T_{eff+}$, and $T_{effBH}$. We will use again the exact
numerical results for the brane and bulk greybody factors, that follow from an analysis
identical to that in the minimal-coupling case -- although the coupling constant $\xi$
modifies the form of the effective potentials that the brane and bulk scalar fields have to
overcome to reach infinity \cite{KPP1}, it has no effect at the asymptotic regimes of the
two horizons; therefore, the asymptotic solutions (\ref{BH-exp}) and (\ref{CO-exp}) as well
as the boundary conditions (\ref{R_BH_num})-(\ref{dR_BH_num}) remain the same.

\begin{figure}[t]
  \begin{center}
\mbox{\includegraphics[width = 0.45 \textwidth] {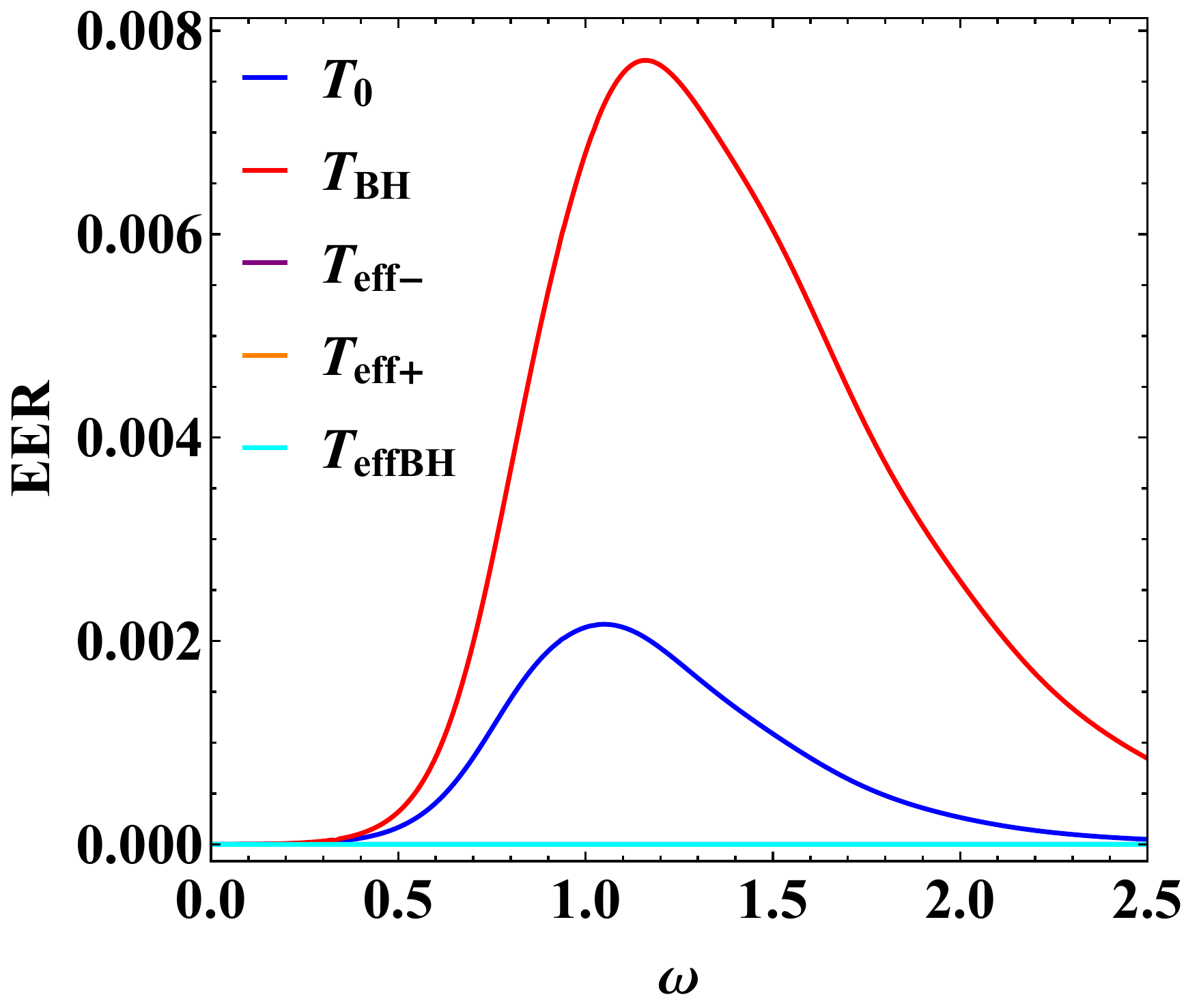}}
\hspace*{0.3cm} {\includegraphics[width = 0.44 \textwidth]
{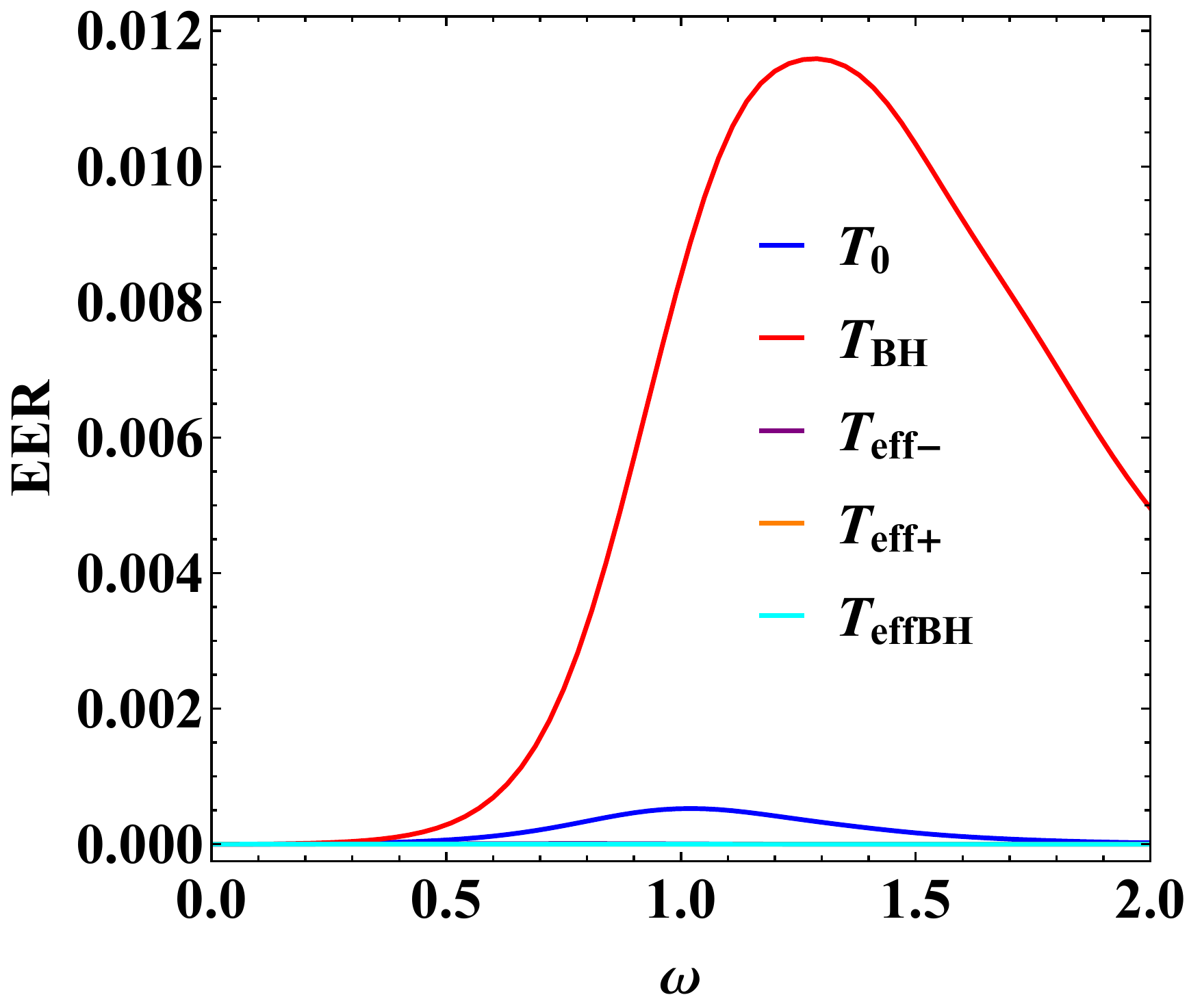}} \hspace*{0.3cm} {\includegraphics[width = 0.45 \textwidth]
{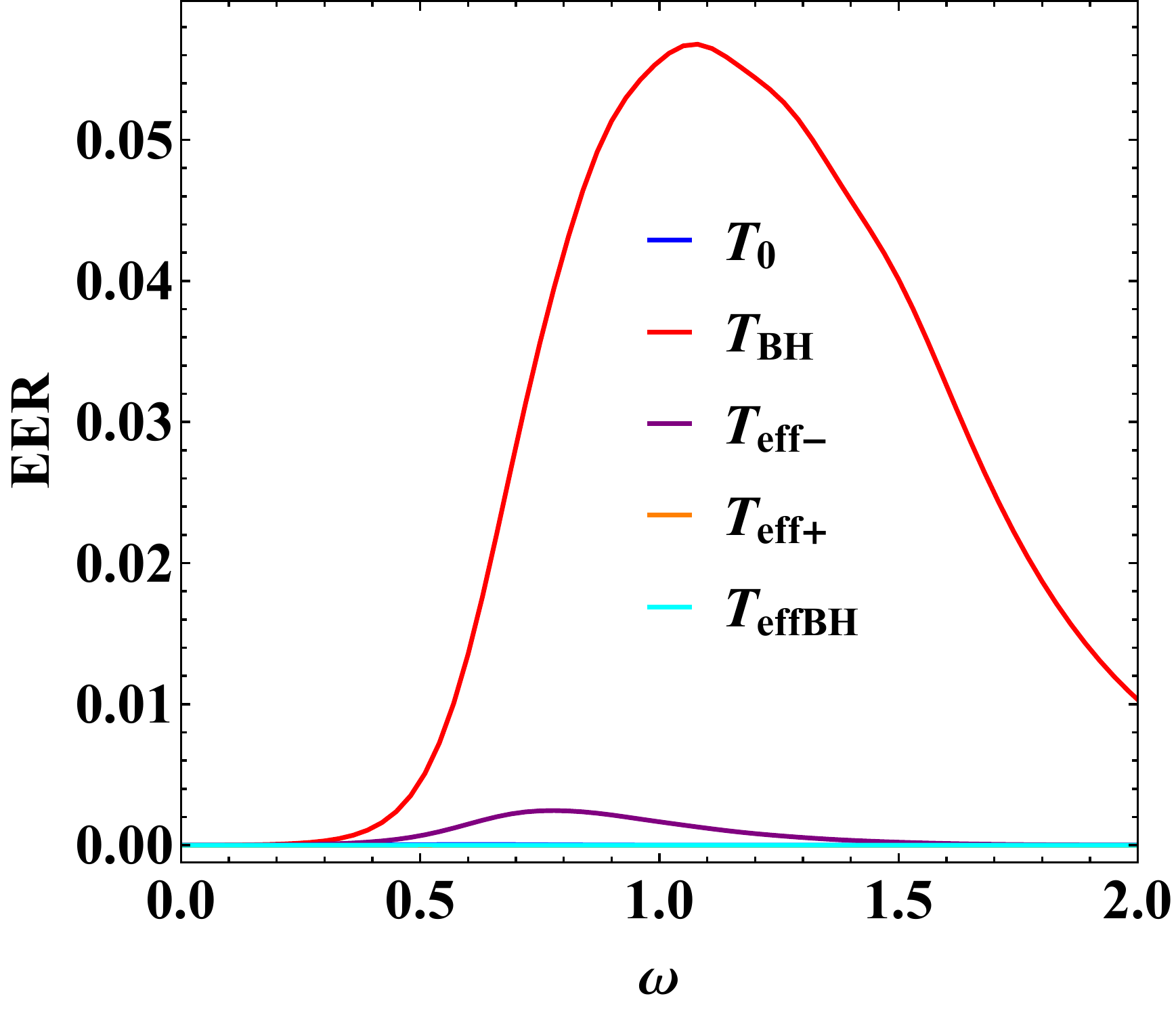}} \hspace*{0.3cm} {\includegraphics[width = 0.45 \textwidth]
{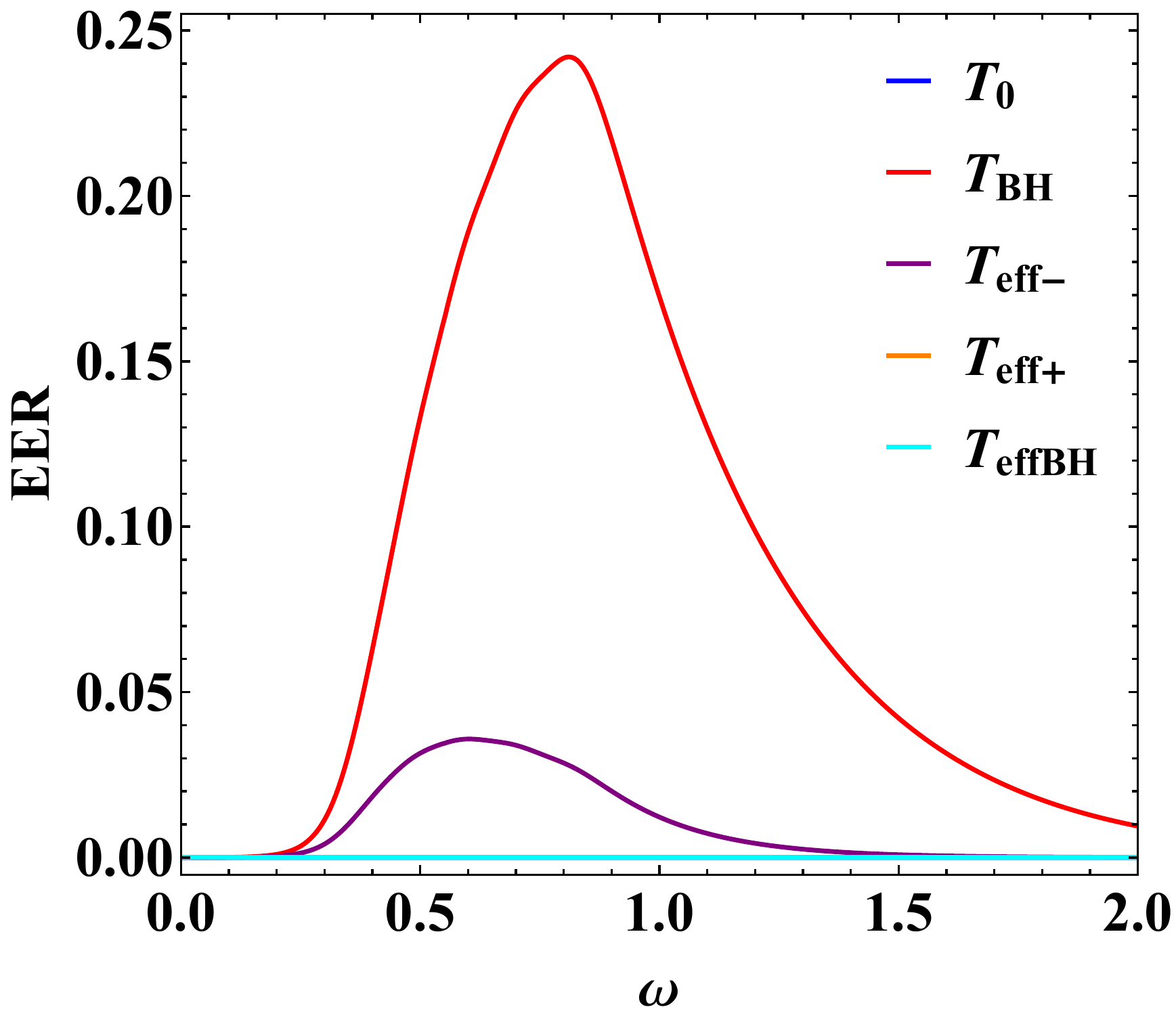}}
    \caption{Energy emission rates for non-minimally coupled brane scalar fields, with $\xi=1$,
from a 6-dimensional ($n=2$) SdS black hole for different temperatures $T$, and for:
    \textbf{(a)} $\Lambda=2$, \textbf{(b)} $\Lambda=2.8$, \textbf{(c)}
    $\Lambda=4$ and \textbf{(d)} $\Lambda=5$ (in units of $r_h^{-2}$).}
   \label{Brane_EER_n2_L_x1}
  \end{center}
\end{figure}

Starting from the emission of non-minimally-coupled scalar fields on the brane, in 
Fig. \ref{Brane_EER_n2_L_x1} we depict the differential energy emission rates for a
6-dimensional SdS black hole, and for the values $\Lambda=2, 2.8, 4$ and 5 of the
bulk cosmological constant. We first note that, in the presence of $\xi$, the emission
curves have returned to their typical shape: as was found in \cite{Crispino, KPP1},
and confirmed also here,  the non-minimal coupling destroys the non-zero asymptotic
limit of the scalar greybody factor in the low-energy limit; as a result, all emission
curves emanate from zero at the low-energy regime. Moreover, the larger the value
of $\xi$, the later in terms of $\omega$ the emission curves rise above the zero value,
in accordance to the effect that the mass of the scalar particle has on the spectra
\cite{Sampaio, KNP1}. Also, by comparing the vertical axes of Figs.
\ref{EER_brane_n2_L}(b,c) and \ref{Brane_EER_n2_L_x1}(a,c), respectively, we observe
that the radiation spectra in the non-minimal case are indeed significantly suppressed,
in accordance to the previous discussion.

\begin{figure}[t]
  \begin{center}
\mbox{\includegraphics[width = 0.45 \textwidth] {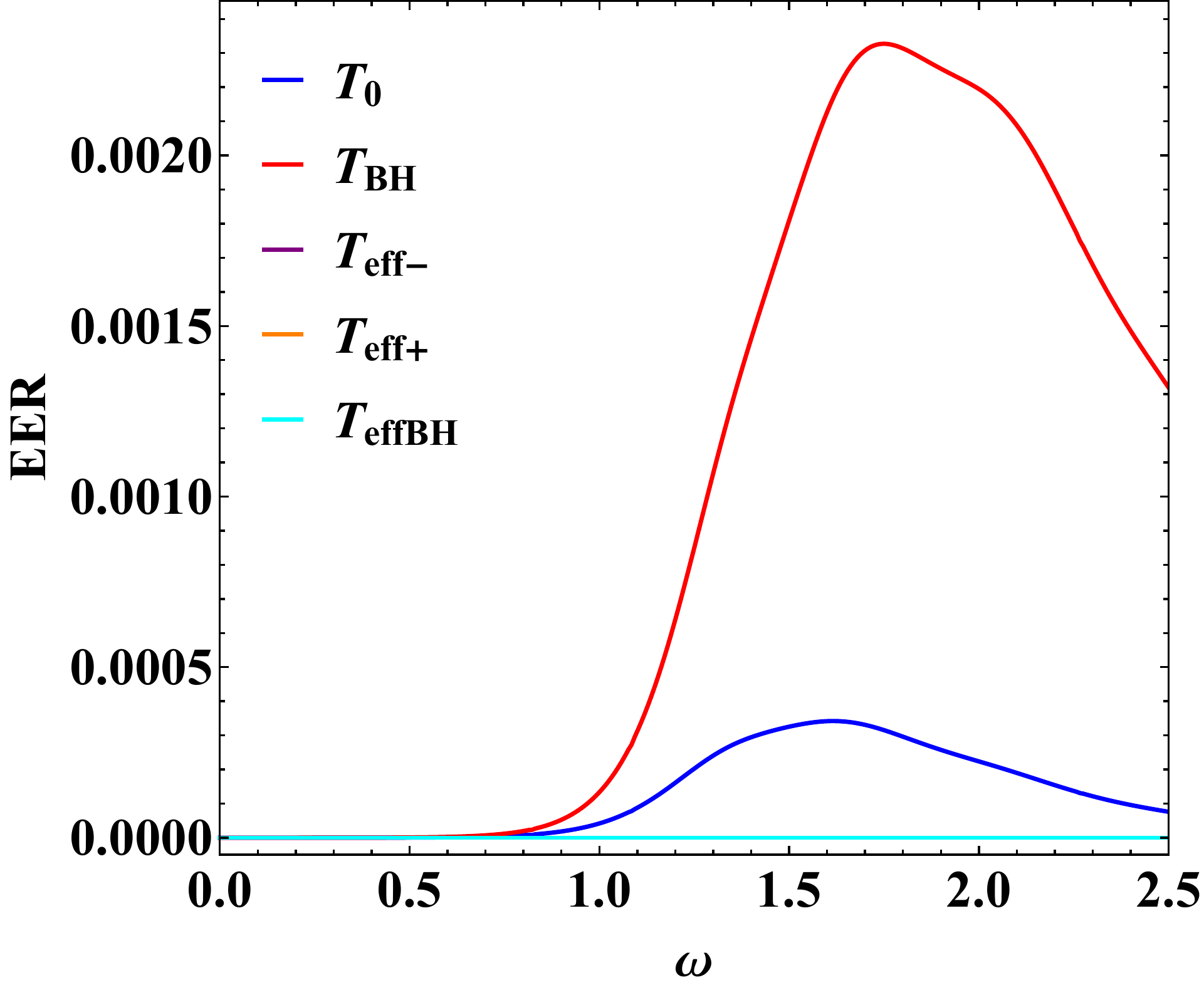}}
\hspace*{0.3cm} {\includegraphics[width = 0.44 \textwidth]
{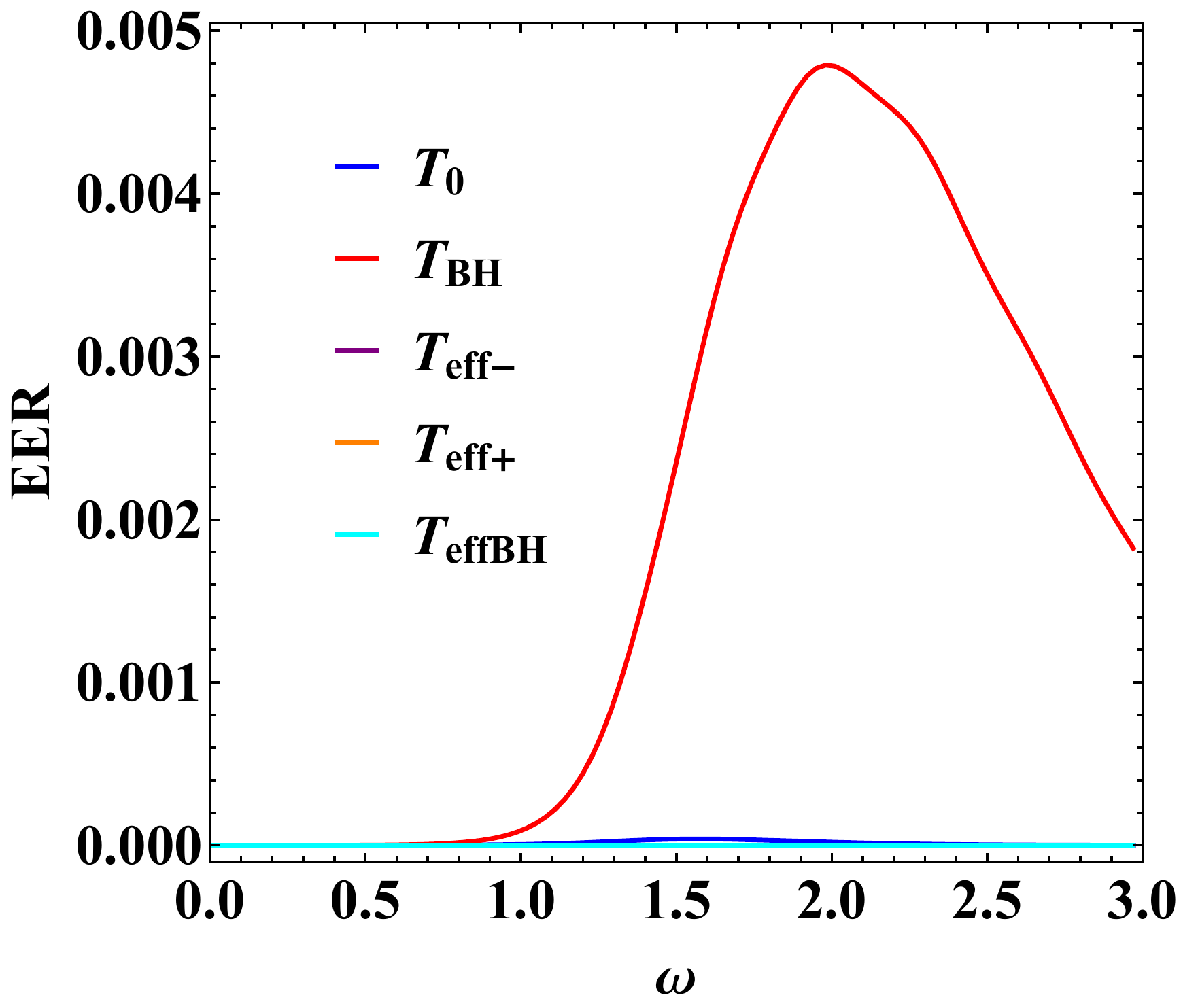}} \hspace*{0.3cm} {\includegraphics[width = 0.45 \textwidth]
{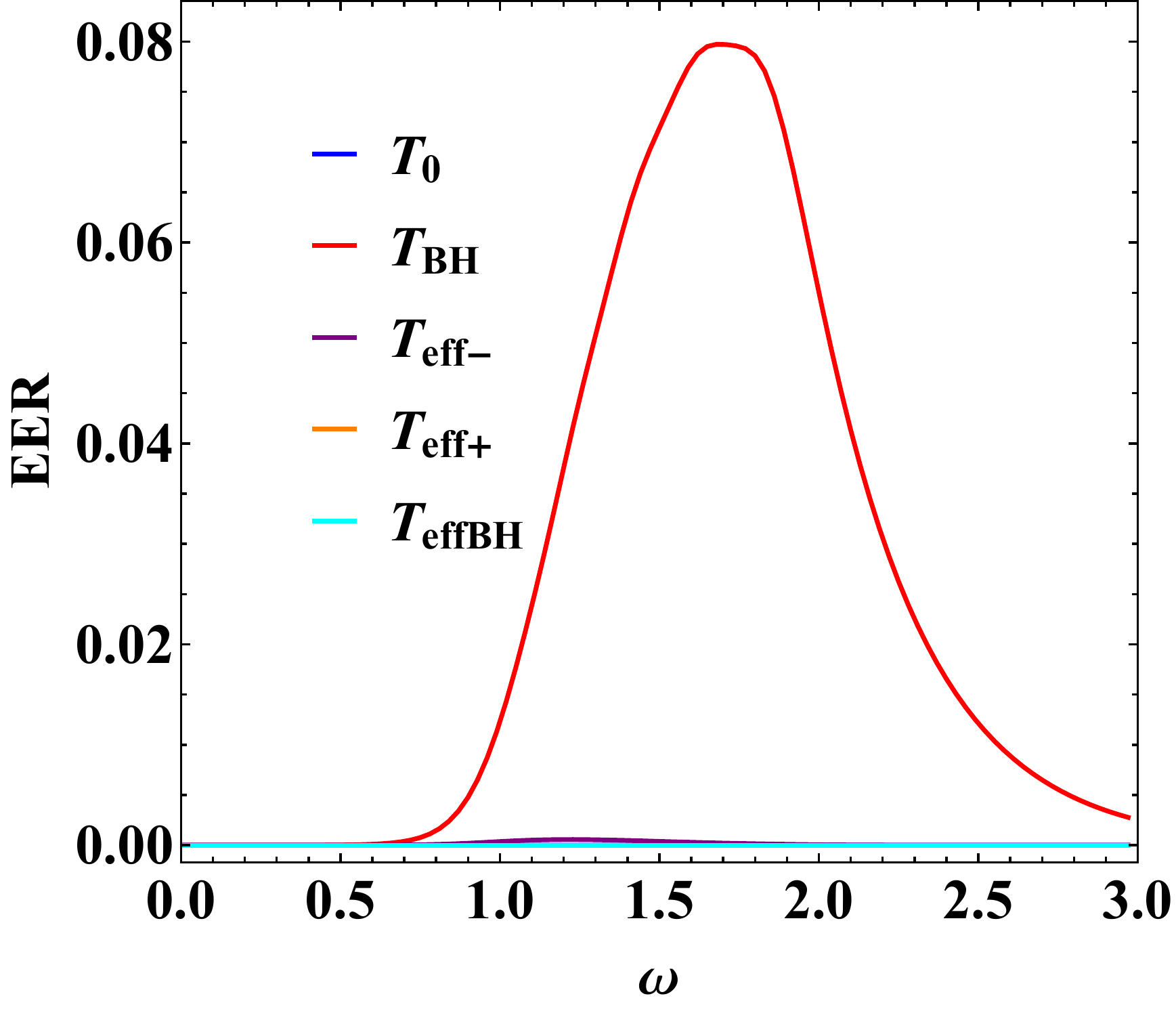}} \hspace*{0.3cm} {\includegraphics[width = 0.45 \textwidth]
{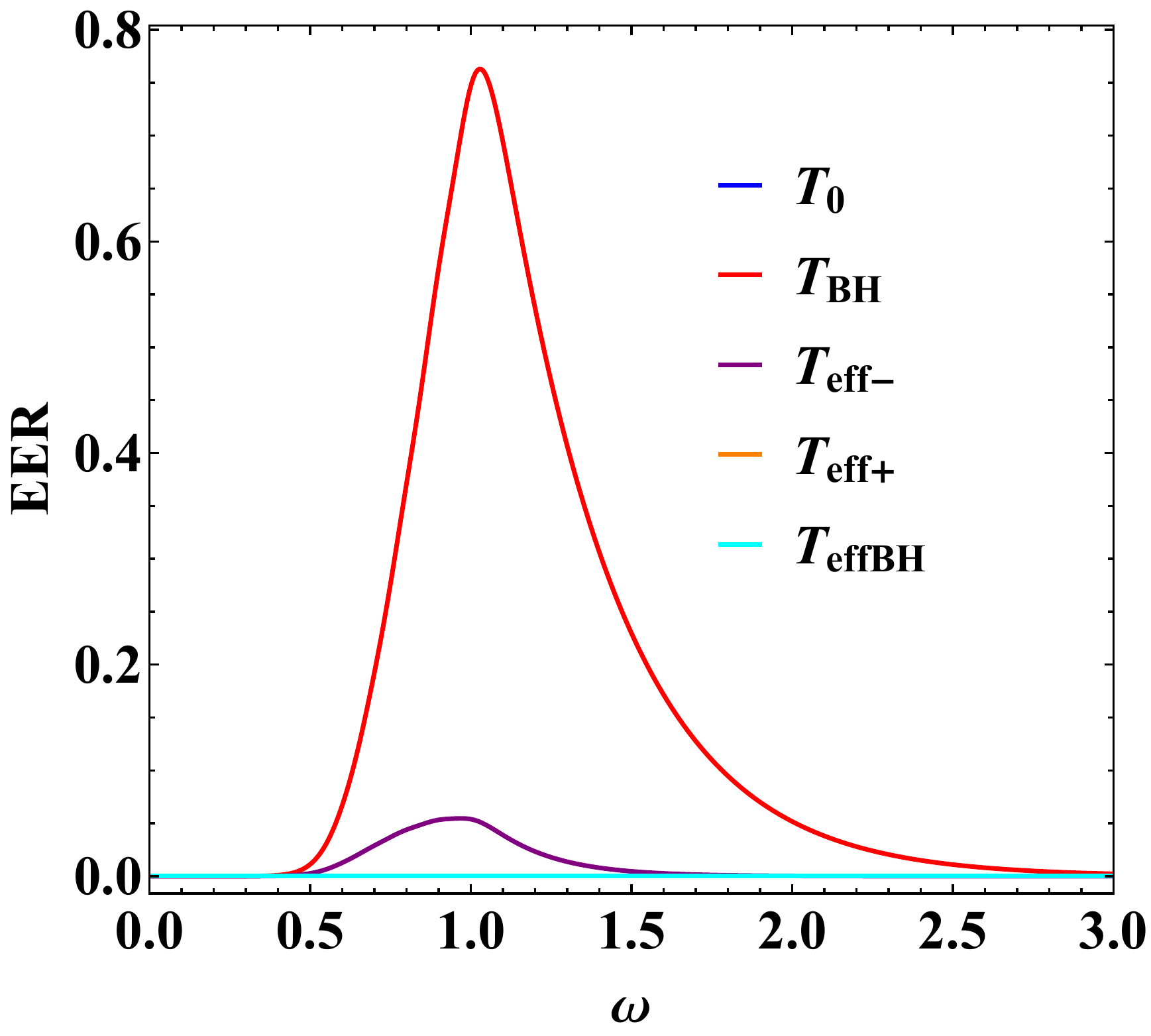}}
    \caption{Energy emission rates for non-minimally coupled bulk scalar fields, with $\xi=1$,
from a 6-dimensional ($n=2$) SdS black hole for different temperatures $T$, and for:
    \textbf{(a)} $\Lambda=2$, \textbf{(b)} $\Lambda=2.8$, \textbf{(c)}
    $\Lambda=4$ and \textbf{(d)} $\Lambda=5$ (in units of $r_h^{-2}$).}
   \label{Bulk_EER_n2_L_x1}
  \end{center}
\end{figure}

This suppression is due to the fact that the greybody factors for both brane and bulk
scalar fields decrease with any increase in the non-minimal coupling constant $\xi$,
and therefore is common to the radiation spectra for the different temperatures. 
As a result, the inclusion of the non-minimal coupling does not modify the general
picture drawn in the previous section. However, some of the radiation spectra are more
sensitive to the changes brought by the presence of the non-minimal coupling. For
example, in Fig. \ref{EER_brane_n2_L} drawn for the minimal-coupling case, we
observe that, for the three effective temperatures
and $T_0$, the maxima of all emission curves are located at the very low-energy limit;
the relatively small magnitude of these temperatures, compared to that of $T_{BH}$,
combined with the enhanced value of the greybody factor for ultra soft particles,
makes the emission of low-energetic particles much more favourable for the black
hole. When the non-minimal coupling is introduced, the emission of soft particles
becomes disfavoured and the radiation spectra for the aforementioned four temperatures
are significantly suppressed. The radiation spectrum for the normalised temperature
$T_{BH}$ is also suppressed, however its relatively large value allows also for the significant
emission of higher-energetic particles and these are not significantly affected by the 
non-minimal coupling. As a result, the relative enhancement of the $T_{BH}$ radiation
spectrum compared to the remaining ones is extended by the non-minimal coupling.
As the critical limit is approached, only the $T_{eff-}$ spectrum manages again to reach
comparable values due to its asymptotic, non-zero value at that regime.

A similar behaviour is observed also in the case where the number of extra dimensions
takes larger values. We have performed the same analysis for $n=5$, and found that
all emission curves for non-minimally coupled brane scalar fields return again to their
typical shape and thus have the emission of low-energy particles suppressed. For small
values of $\Lambda$, and due to the enhancement with $n$ that characterizes both $T_0$
and $T_{BH}$ (see Fig. \ref{Temp_n_L08}) the difference in the corresponding two radiation
spectra is smaller compared to the case with $n=2$; as $\Lambda$ however
increases, the $T_0$ radiation spectrum is constantly suppressed reaching a negligible
value at the critical limit. Of the effective temperature, only $T_{eff-}$ manages to 
support a relatively significant spectrum and that is realised very close to the critical limit.

We now turn to the case of the emission of non-minimally-coupled scalar fields emitted
in the bulk. The radiation spectra for the different temperatures and for the case with $n=2$
are now depicted in Fig. \ref{Bulk_EER_n2_L_x1}, again for the value $\xi=1$ and for the
same four values of the cosmological constant. A similar picture emerges also here: the
$T_0$ radiation spectrum is significant only in the low $\Lambda$ regime, the $T_{eff-}$
becomes important near the critical limit, while the other two radiation spectra for $T_{eff+}$
and $T_{effBH}$ fail to acquire any significant value at any $\Lambda$ regime. The 
radiation spectrum for $T_{BH}$ is the one that dominates over the whole energy regime
and for the entire $\Lambda$ range. The same behaviour is observed also for $n=5$.

Let us finally note that the dominance of the bulk emission channel in the large
$\Lambda$ regime \cite{KPP3} is confirmed also in the case of non-minimal coupling
and even for models with a small number of extra dimensions. As the comparison of the
vertical axes of Figs. \ref{Brane_EER_n2_L_x1} and \ref{Bulk_EER_n2_L_x1} reveals, the
differential energy emission rate in the bulk exceeds that on the brane as soon as
$\Lambda$ becomes approximately larger than 3, and stays dominant for the
remaining half of the allowed range. 


\section{Bulk-over-Brane Relative Emissivities}

A final question that we would like to address in this section is that of the effect of the different
temperatures on the total emissivities in the bulk and on the brane, and more particularly on the
bulk-over-brane emissivity ratio. In our previous work \cite{KPP3}, we calculated the
total power emitted by the SdS black hole over the whole frequency range in both the brane and
bulk channels, by employing the Bousso-Hawking $T_{BH}$ normalization for the temperature.
Here, we generalise this analysis to cover all five temperatures $T_0$, $T_{BH}$, $T_{eff-}$,
$T_{eff+}$ and $T_{effBH}$, and compare the corresponding results. We also extend our previous
study by considering the whole range of values for the bulk cosmological constant, from a 
vanishing value up to its critical limit.

The quantity of interest, namely the ratio of the total power emitted in the bulk over the corresponding
total power on the brane, for the case with $n=2$ and for four different values of the coupling
constant $\xi$, i.e. $\xi=0,0.5,1,2$, is presented in Tables 1 through 4. The five columns of each
Table give the total ratio for five values of the cosmological constant that span the entire allowed
range, i.e. for $\Lambda=0.3,1,2,4,5$. Let us see first how the change in the value of $\Lambda$
affects our results. For small values of $\Lambda$, and independently of the value of $\xi$, the
brane emission channel clearly dominates over the bulk one; however, as $\Lambda$ increases,
the bulk emission channel gradually becomes more and more important. This is due to the
fact that for an increasing cosmological constant the bulk emission curves move to the right,
thus allowing for the emission of a larger number of high-energetic particles compared
to that on the brane, but also the maximum height of the bulk curves soon overpasses the
one of the brane curves by a factor of 3. For the $T_{BH}$ and $T_{eff-}$ temperatures, that
retain a significant value near the critical limit, the bulk-over-brane ratio well exceeds unity 
thus rendering the bulk channel the dominant one in the emission process of the black hole -
the tendency of $T_{BH}$ to overturn the power ratio in favour of the bulk channel was already
anticipated by the results of \cite{KPP3}. The only exception to the above behaviour is the
one exhibited by the bare temperature $T_0$: the enhancement of the bulk-over-brane
ratio with $\Lambda$ is observed only in the case of minimal coupling whereas this ratio
decreases for all values $\xi \neq 0$, as $\Lambda$ increases towards its critical value.
We may interpret this as the result of the disappearance of the low-energy modes as soon
as the coupling constant $\xi$ takes a non-vanishing value: the emission curves for $T_0$
have their maxima at the low-energy regime and are thus mostly affected when these are
banned from the emission spectrum -- according to our results, this change affects more
the bulk channel rather than the brane one causing the suppression of the bulk-over-brane
ratio. 


\begin{table}[t!]
\caption{Bulk over brane total emissivity for $n=2$ and $\xi=0$} 
\centering 
\begin{tabular}{|c || c| c| c| c| c|} 
\hline\ 
$  \Lambda \rightarrow$ & 0.3 & 1 & 2 & 4 & 5 \\ [0.5ex] 
\hline 
\hline $T_0$& 0.259268  & 0.304247 & 0.402190  & 0.663547 & 0.781833  \\ 
\hline$T_{BH}$& 0.338245 & 0.506324 & 0.798603 & 1.929660 & 3.247190  \\
\hline$T_{eff-}$& 0.032997 & 0.132329 & 0.319508 & 0.860880 & 2.071590  \\ 
\hline$T_{eff+}$& 0.032507 & 0.125599 & 0.298895 & 0.717772 & 0.884068  \\ 
\hline$T_{effBH}$& 0.032950 & 0.130510 & 0.309000 & 0.669669 & 0.792598  \\ 
\hline 
\end{tabular}
\label{table n2xi0} 
\end{table}

\begin{table}[t!]
\caption{Bulk over brane total emissivity for $n=2$ and $\xi=0.5$} 
\centering 
\begin{tabular}{|c || c| c| c| c| c|} 
\hline\ 
$  \Lambda \rightarrow$ & 0.3 & 1 & 2  & 4 & 5 \\ [0.5ex] 
\hline 
\hline $T_0$& 0.281627  & 0.220836 & 0.160691 &  0.089933 & 0.067954  \\ 
\hline$T_{BH}$& 0.369359  & 0.450873 & 0.629061 & 1.617200 & 2.962410  \\
\hline$T_{eff-}$& 0.003762  & 0.012441 & 0.038311 & 0.432708 & 1.710000  \\ 
\hline$T_{eff+}$& 0.003424  & 0.008841 & 0.014009 & 0.019979 & 0.021436  \\ 
\hline$T_{effBH}$& 0.003725  & 0.011167 & 0.022578 & 0.046074 & 0.052124  \\ 
\hline 
\end{tabular}
\label{table n2xi05} 
\end{table}

\begin{table}[t!]
\caption{Bulk over brane total emissivity for $n=2$ and $\xi=1$} 
\centering 
\begin{tabular}{|c || c| c| c|  c| c|} 
\hline\ 
$  \Lambda \rightarrow$ & 0.3 & 1 & 2 & 4 & 5 \\ [0.5ex] 
\hline 
\hline $T_0$& 0.286455  & 0.165240 & 0.089413  & 0.032550& 0.020609  \\ 
\hline$T_{BH}$& 0.380420 & 0.387464 & 0.500779  & 1.364060 & 2.704060  \\
\hline$T_{eff-}$& 0.001233  & 0.003214 & 0.011410 &  0.279735 & 1.433260  \\ 
\hline$T_{eff+}$& 0.001140  & 0.002529 & 0.003787&  0.005227 & 0.005582  \\ 
\hline$T_{effBH}$& 0.001222  & 0.002907 & 0.005497 &  0.012099 & 0.013918  \\ 
\hline 
\end{tabular}
\label{table n2xi1} 
\end{table}

\begin{table}[t!]
\caption{Bulk over brane total emissivity for $n=2$ and $\xi=2$} 
\centering 
\begin{tabular}{|c || c| c| c| c| c|} 
\hline\ 
$  \Lambda \rightarrow$ & 0.3  & 1 & 2  & 4 & 5 \\ [0.5ex] 
\hline 
\hline $T_0$& 0.280978 & 0.099559 & 0.035998 &  0.007446 & 0.003698  \\ 
\hline$T_{BH}$& 0.382963  & 0.287373 & 0.331984  & 1.002190 & 2.289020  \\
\hline$T_{eff-}$& 0.000222 & 0.000471 & 0.001935 &  0.138896 & 1.045890 \\ 
\hline$T_{eff+}$& 0.000216 & 0.000410 & 0.000580& 0.000778 & 0.000828  \\ 
\hline$T_{effBH}$& 0.000221  & 0.000438 & 0.000738 & 0.001767 & 0.002089  \\ 
\hline 
\end{tabular}
\label{table n2xi2} 
\end{table}


If we now turn our attention to the role of the non-minimal coupling constant $\xi$ in
the value of the bulk-over-brane ratio, we find that the overall behaviour is a suppression
of this quantity as $\xi$ increases. This behaviour holds for almost all values of the
cosmological constant apart from the lower part of its allowed regime where, in contrast,
the bulk-over-brane ratio exhibits an enhancement without however exceeding unity. On
the other hand, despite the suppression with $\xi$, the bulk-over-brane ratio retains 
values above unity when $\Lambda$ tends  to its critical limit.


\begin{table}[t!]
\caption{Bulk over brane total emissivity for $n=5$ and $\xi=0$} 
\centering 
\begin{tabular}{|c || c| c| c| c| c|} 
\hline\ 
$  \Lambda \rightarrow$  & 1 & 4 & 10 & 13 & 18 \\ [0.5ex] 
\hline 
\hline $T_0$ & 0.296070 & 0.299653 & 0.357216 & 0.422606 & 0.584868  \\ 
\hline$T_{BH}$ & 0.419245 & 0.818056 & 2.578580 & 4.629670 & 14.18230  \\
\hline$T_{eff-}$ & 0.000267 & 0.010603 & 0.140588 & 0.328066 & 4.192670  \\ 
\hline$T_{eff+}$ & 0.000265 & 0.010319 & 0.137045 & 0.291825 & 0.658816  \\ 
\hline$T_{effBH}$ & 0.000267 & 0.010549 & 0.134856 & 0.273098 & 0.559205  \\ 
\hline 
\end{tabular}
\label{table n5xi0} 
\end{table}

\begin{table}[ht!]
\caption{Bulk over brane total emissivity for $n=5$ and $\xi=0.5$} 
\centering 
\begin{tabular}{|c || c| c| c| c| c|} 
\hline\ 
$  \Lambda \rightarrow$ & 1 & 4 & 10 & 13 & 18 \\ [0.5ex] 
\hline 
\hline $T_0$ & 0.468836 & 0.288097 & 0.099659 & 0.054591 & 0.016835  \\ 
\hline$T_{BH}$ & 0.641474 & 0.841435 & 1.770690 & 3.060490 & 11.19970  \\
\hline$T_{eff-}$ & 3.152 $10^{(-6)}$ & 0.000090 & 0.002028 & 0.018275 & 2.231760  \\ 
\hline$T_{eff+}$& 2.898 $10^{(-6)}$ & 0.000071 & 0.000552 & 0.000923 & 0.001500  \\ 
\hline$T_{effBH}$& 3.139 $10^{(-6)}$ & 0.000086 & 0.000938 & 0.002127 & 0.005982  \\ 
\hline 
\end{tabular}
\label{table n5xi05} 
\end{table}

\begin{table}[ht!]
\caption{Bulk over brane total emissivity for $n=5$ and $\xi=1$} 
\centering 
\begin{tabular}{|c || c| c| c| c| c|} 
\hline\ 
$  \Lambda \rightarrow$  & 1 & 4 & 10 & 13 & 18 \\ [0.5ex] 
\hline 
\hline $T_0$ & 0.664875 & 0.248447 & 0.040067 & 0.015499 & 0.002610  \\ 
\hline$T_{BH}$ & 0.890165 & 0.778299 & 1.195190 & 2.049140 & 9.026680  \\
\hline$T_{eff-}$ & 3.679 $10^{(-7)}$ & 0.000007 & 0.000200 & 0.003575 & 1.293840  \\ 
\hline$T_{eff+}$ & 3.956 $10^{(-7)}$ & 0.000006 & 0.000054 & 0.000095 & 0.000164  \\ 
\hline$T_{effBH}$ & 3.683 $10^{(-7)}$ & 0.000007 & 0.000080 & 0.000187 & 0.000616  \\ 
\hline 
\end{tabular}
\label{table n5xi1} 
\end{table}

\begin{table}[ht!]
\caption{Bulk over brane total emissivity for $n=5$ and $\xi=2$} 
\centering 
\begin{tabular}{|c || c| c| c| c| c|} 
\hline\ 
$  \Lambda \rightarrow$  & 1 & 4 & 10 & 13 & 18 \\ [0.5ex] 
\hline 
\hline $T_0$ & 1.162700 & 0.179527 & 0.009087 & 0.002170 & 0.000160  \\ 
\hline$T_{BH}$ & 1.509360 & 0.632852 & 0.585960 & 1.010350 & 6.207500  \\
\hline$T_{eff-}$ & 0.000274 &1.054 $10^{(-6)}$ & 6.508 $10^{(-6)}$ & 0.000305 & 0.514108  \\ 
\hline$T_{eff+}$ & 0.000299 & 1.653 $10^{(-6)}$ & 1.827 $10^{(-6)}$ & 3.402 $10^{(-6)}$ & 6.292 $10^{(-6)}$  \\ 
\hline$T_{effBH}$ & 0.000275 &1.033 $10^{(-6)}$ & 2.328 $10^{(-6)}$ & 5.573 $10^{(-6)}$ & 0.000022  \\ 
\hline 
\end{tabular}
\label{table n5xi2} 
\end{table}


When we increase the number of the extra dimensions, all the above effects become
amplified. In Tables 5 through 8, we display the value of the bulk-over-brane ratio
for the case with $n=5$,  for the same four values of the non-minimal coupling
constant $\xi$  and for five indicative values of the bulk cosmological constant,
i.e. $\Lambda=1,4,10,13$ and $18$, that again span the entire allowed regime.
The dominance of the bulk channel over the brane one for $T_{BH}$ and $T_{eff-}$,
as $\Lambda$ approaches its critical limit, is now much more prominent
with the overall energy emitted in the bulk surpassing the one emitted on the brane
by a factor of even larger than 10. The suppression of the energy ratio as $\xi$ increases
is also obvious here, but again this suppression does not prevent the bulk from
becoming the dominant channel at the critical limit. What is different in this case
from the $n=2$ case is that the enhancement with $\xi$ for small values of the cosmological
constant, noted also in the case with $n=2$, is now adequate to cause the dominance
of the bulk channel over the brane one for the bare $T_0$ and normalised $T_{BH}$
temperatures - for the latter temperature, this effect was also observed in \cite{KPP3}.


\section{Conclusions}
Over the years, the study of the thermodynamics of the Schwarzschild-de Sitter spacetime
has proven to be a challenging task. The existence of two different horizons, the
black-hole and the cosmological one -- each with its own temperature expressed
in terms of its surface gravity -- results into the absence of a true
thermodynamical equilibrium. On the other hand, the absence of an asymptotically-flat
limit led to the formulation of a normalised temperature for the black hole \cite{BH}
more that two decades ago. Both problems become more severe in the limit of large
cosmological constant when the two horizons are located so close to each other that
the argument of the two independent thermodynamics, valid at the two horizons,
comes into question. As a result, the notion of the effective temperature of the SdS
spacetime was proposed \cite{Urano, Lahiri, Bhatta, LiMa} that implements both the
black-hole and the cosmological horizon temperatures.

In the context of the present work, we have focused on the case of the higher-dimensional
Schwarzschild-de Sitter black hole, and have formed a set of five different temperatures:
the bare black-hole temperature $T_0$, based on its surface gravity, the normalised
black-hole temperature $T_{BH}$ and three effective temperatures for the SdS spacetime,
$T_{eff-}$, $T_{eff+}$ and $T_{effBH}$ -- the latter three are inspired by four-dimensional
analyses, where the cosmological constant plays the role of the pressure of the system,
and are combinations of the black-hole and cosmological horizon temperatures.
We have first studied the dependence of the aforementioned temperatures on the
value of the cosmological constant, as this is varied from zero to its maximum
allowed value, set by the critical limit where the two horizons coincide. In the limit
of vanishing cosmological constant, the black-hole temperatures $T_0$ and $T_{BH}$
reduce to the temperature of an asymptotically-flat, higher-dimensional Schwarzschild
black hole as expected; on the other hand, all three effective temperatures tend
to zero, an artificially ill behaviour due to the fact that $\Lambda$ (or, equivalently
the pressure of the system) is not allowed to vanish. In the opposite limit, that of the
critical value, it is the normalised $T_{BH}$ and effective $T_{eff-}$ temperatures that
have a common behaviour reaching a non-vanishing asymptotic value; the other
three temperatures all vanish in the same limit. We then examined the dependence
of the temperatures on the number of extra dimensions. Here, the five temperatures
were found to fall again into two categories: the black-hole temperatures $T_0$ and
$T_{BH}$ both are enhanced with $n$ while all effective temperatures predominantly
are suppressed. Overall, the normalised $T_{BH}$ temperature was found to be the
dominant one for all values of $\Lambda$ and $n$. 

The set of five temperatures was then used to derive the Hawking radiation spectra
for a free, massless scalar field propagating both on the brane and in the bulk.
We considered the cases where the number of extra dimensions had a small ($n=2$)
and a large ($n=5$) value: in each case, we chose four different values for the
cosmological constant that covered the allowed regime from zero to the critical
value. For both brane and bulk radiation spectra, the emission curves closely
followed the behaviour of the temperatures: for small $\Lambda$, the emission
curves for all effective temperatures were significantly suppressed while the ones
for the black-hole temperatures were the dominant ones. As $\Lambda$ increased,
the emission rate for the bare $T_0$ started to become suppressed while the one
for the effective $T_{eff-}$ started to become important. Near the critical limit,
it is the two temperatures, $T_{BH}$ and $T_{eff-}$, with the non-vanishing values
that lead to the dominant emission curves. It is worth noting that the two
effective temperatures $T_{eff+}$ and $T_{effBH}$ support a non-negligible
emission rate only for intermediate values of the cosmological constant, where
they favour the emission of very low-energetic scalar particles. The emission rate
for the normalised temperature $T_{BH}$ is the one that constantly rises
as $\Lambda$ gradually increases, being clearly the dominant one: for $n=2$,
the peak of the emission curve on the brane for
$T_{BH}$ rises to a height that is 2 times larger than that for $T_0$ at
the low $\Lambda$-regime and 5 times larger than that for $T_{eff-}$
at the high $\Lambda$-regime; these factors increase even more as $n$
increases, or when we study the bulk emission channel.

For the case of a minimally-coupled scalar field, all emission curves were found
to have non-zero asymptotic values at the very-low part of the spectrum due
to the well-known behaviour of the greybody factor both on the brane and in
the bulk. As a result, a significant number of soft particles are expected to be
emitted; in fact, for the three effective temperatures $T_{eff-}$, $T_{eff+}$,
$T_{effBH}$ (for small values of values of $\Lambda$) and for $T_{0}$
(for large
values of $\Lambda$) this is where the peak of the emission curves is located.
When the non-minimal coupling to the scalar curvature is turned on, the 
emission curves for all five temperatures resume their usual shape. The
general behaviour regarding the comparative strength of the emission curves
for the different temperatures observed in the case of the minimal coupling holds
also here. The emission curve for the normalised temperature $T_{BH}$ is again
the dominant one over the entire $\Lambda$-regime, with only the emission curves
for $T_0$ and $T_{eff-}$ reaching significant values at low and large values
of $\Lambda$, respectively.

The exact analysis performed in the context of this work serves not only as
a comparison of the radiation spectra, that follow by using different 
temperatures for the Schwarzschild-de Sitter spacetime, but also as a source
of information regarding their behaviour as the cosmological constant 
varies from a very small value to the largest allowed one at the critical limit.
The complete radiation spectra reveal that as $\Lambda$ increases, the emission
of energy from the black hole along the brane and bulk channels very quickly become
comparable, and even for low values of the number of extra dimensions, the bulk
emission eventually dominates over the brane one. The exact total emissivities that
were calculated in Section 5 demonstrated exactly this effect: apart from the case
of $T_0$ when $\xi \neq 0$, the bulk-over-brane ratio exhibits a significant enhancement
as $\Lambda$ increases and, in fact, renders the bulk channel the dominant emission
channel of the SdS black hole for the temperatures $T_{BH}$ and $T_{eff-}$, i.e.
for the temperatures that retain a non-vanishing value near the critical limit. In
addition, when the number of extra dimensions is large enough, the bulk was
found to dominate over the brane even for values of $\Lambda$ much lower
than its critical limit as long as the value of the non-minimal coupling constant
$\xi$ was large enough; in this case, the bulk dominance was obtained also
for the bare temperature $T_0$. 

In conclusion, choosing a particular form for the temperature of an SdS black
hole, i.e. the bare, the normalised or an effective one, plays a paramount role 
in the form of the obtained radiation spectra. Some of the suggested temperatures
fail even to produce a significant emission rate, others lead to an emission only
for very small or very large values of the bulk cosmological constant. Our results
clearly reveal that the normalised temperature $T_{BH}$, the one that makes amends
for the absence of an asymptotically-flat limit in a Scwarzschild-de Sitter spacetime,
is the one that produces the most robust radiation spectra over the entire regime
of the bulk cosmological constant.

{\bf Acknowledgement} T.P. would like to thank the Alexander S. Onassis Public Benefit Foundation for financial support.


\end{document}